\documentclass[ALICE,manyauthors]{cernphprep}

\usepackage{graphicx}
\usepackage{hyperref}   %% produces clickable links in xdvi
\usepackage{flafter}
\usepackage{amsmath}
\usepackage{float}
\usepackage{wrapfig}
\usepackage{epsfig}
\usepackage{epstopdf}
\usepackage{mathtools}
\usepackage{amssymb}
\usepackage{multicol}
\usepackage{multirow}
\usepackage{color}
\usepackage[modulo]{lineno}
\usepackage{array}
\usepackage{soul}
\usepackage{caption}
\usepackage{enumitem}
\usepackage[titletoc]{appendix}
\usepackage{booktabs}
\usepackage[usenames,dvipsnames]{xcolor}
\usepackage{cite}
\usepackage{bm}
\usepackage{svn-multi}

\svnid{$Id: jetmass_paper.tex 3049 2017-02-02 19:06:08Z loizides $}

\DeclareBoldMathCommand\boldlangle{\left\langle}
\DeclareBoldMathCommand\boldrangle{\right\rangle}

\newcommand{\rom}[1]{{\mathrm{#1}}}   % switch to rom in math 

\newcommand{\kt}{{$k_{\rm T}$}}

\newcommand{\PbPb}{\mbox{Pb--Pb}}
\newcommand{\pp}{pp}
\newcommand{\pPb}{\mbox{p--Pb}}
\newcommand{\sNN}{\ensuremath{\sqrt{s_\mathrm{NN}}}}
\newcommand{\GeV}{GeV}
\newcommand{\GeVc}{\text{\GeV/}\ensuremath{c}}
\newcommand{\GeVcsq}{\text{\GeV/}\ensuremath{c^{2}}}

\newcommand{\pt}{\ensuremath{p_\rom{T}}}

\newcommand{\deltapt}{\ensuremath{\rom{\delta}\pt}}

\newcommand{\ptjetrawch}{\ensuremath{p_\rom{T,ch\;jet}^\rom{raw}}}
\newcommand{\ptjetsubch}{\ensuremath{p_\rom{T,ch\;jet}^\rom{sub}}}
\newcommand{\ptjetch}{\ensuremath{p_\rom{T,ch\;jet}}}

\newcommand{\mjetch}{\ensuremath{M_\rom{ch\;jet}}}
\newcommand{\deltaM}{\ensuremath{\rom{\delta}M}}

\newcommand{\RAA}{\ensuremath{R_{\rom{AA}}}}

\newlength{\myfigwidth}
\begin{document}

%%%%%%%%%%%%%%%  Title page %%%%%%%%%%%%%%%%%%%%%%%%
%
\begin{titlepage}
\PHyear{2017}           % % required, will be obtained from PH
\PHnumber{016}             % % required, will be obtained from PH
\PHdate{25 Jan}  
%

%%% Put your own title + short title here:
\title{First measurement of jet mass in \PbPb{} and \pPb{} collisions at the LHC} % \newline \newline \version}
\ShortTitle{First measurement of jet mass in \PbPb{} and \pPb{} collisions}   % appears on right page headers
%
%%% Do not change the next lines!
\Collaboration{ALICE Collaboration%
         \thanks{See Appendix~\ref{app:collab} for the list of collaboration
                      members}}
\ShortAuthor{ALICE Collaboration}      % appears on left page headers, do not change
%
%\linenumbers

\begin{abstract}
This letter presents the first measurement of jet mass in \PbPb{} and \pPb{} collisions at $\sNN = 2.76$~TeV and $\sNN = 5.02 $~TeV, respectively. Both the jet energy and the jet mass are expected to be sensitive to jet quenching in the hot Quantum Chromodynamics (QCD) matter created in nuclear collisions at collider energies. Jets are reconstructed from charged particles using the anti-\kt{} jet algorithm and resolution parameter $R = 0.4$. The jets are measured in the pseudorapidity range $|\eta_{\rom{jet}}|<0.5$ and in three intervals of transverse momentum between 60 \GeVc{} and 120 \GeVc{}. The measurement of the jet mass in central \PbPb{} collisions is compared to the jet mass as measured in \pPb{} reference collisions, to vacuum event generators, and to models including jet quenching. It is observed that the jet mass in central \PbPb{} collisions is consistent within uncertainties with \pPb{} reference measurements. Furthermore, the measured jet mass in \PbPb{} collisions is not reproduced by the quenching models considered in this letter and is found to be consistent with PYTHIA expectations within systematic uncertainties.

\end{abstract}
\end{titlepage}
\setcounter{page}{2}

%\linenumbers
\section{Introduction}\label{sec:intro}
This letter presents the first measurement of jet mass in \PbPb{} and \pPb{} collisions at $\sNN = 2.76$ TeV and $\sNN = 5.02 $ TeV, respectively. Both the jet energy and the jet mass are expected to be sensitive to jet quenching in the hot Quantum Chromodynamics (QCD) matter, the Quark-Gluon-Plasma (QGP), created in ultra-relativistic nuclear collisions. % at collider energies. 
Scattering processes with large momentum transfer, $Q^{2}$, between the quarks and the gluons (partons) constituents of colliding nucleons occur early in the collision (at a time $< 1\,\text{fm}/c$). Outgoing partons carry a net color charge and evolve from high to low virtuality producing parton showers, which eventually hadronize into collimated sprays of particles, called jets. Interactions of the outgoing partons with the hot and dense QGP created in heavy-ion collisions may modify the angular and momentum distributions of hadronic jet fragments relative to jets fragmenting in vacuum. This process, known as jet quenching, can be used to probe the properties of the hot QCD medium \cite{Bjorken82thy,Appel,Gyulassy:1990ye,Baier:1994bd}. 

Jet quenching has been investigated at the Relativistic Heavy Ion Collider (RHIC) \cite{Adcox:2001jp,Adler:2003qi,Adams:2003kv,Arsene:2003yk,Back:2003qr} and at the Large Hadron Collider (LHC) 
\cite{Aamodt:2010jd,Aamodt:2011vg,Aad:2015wga,CMS:2012aa,Aad:2010bu,Chatrchyan:2012nia,Aad:2012vca,Abelev:2013kqa,Adam:2015ewa,Chatrchyan:2014ava,Chatrchyan:2013kwa} via measurements of high-$\pt$ hadrons and fully reconstructed jets in nucleus--nucleus (AA) collisions and pp (vacuum) collisions. These measurements have shown a suppression of hadron and jet yields in AA collisions and modest modifications of the longitudinal fragment distribution and the radial profile of jets relative to jets produced in pp collisions within the typical jet cone of $0.3-0.4$ at the LHC.
%The fragmentation process of a high energetic parton propagating through a dense medium, created in a heavy-ion collisions, is potentially modified compared to the parton shower in vacuum. Collimated jets have a small jet mass while jets with a broad profile have a larger jet mass.  
The jet mass is sensitive to the initial virtuality of the parton at the origin of the shower \cite{Majumder:2014gda}. Energy-momentum exchange with the hot QCD medium may temporarily increase the virtuality of the propagating partons, leading to a larger gluon radiation probability \cite{Vitev:2005yg,Renk:2009nz,Renk:2010zx,Armesto:2011ht}. This would result in a broadening of the jet profile and an increase of the jet mass, if a significant amount of the radiated gluons are captured within the jet cone used for reconstruction. However, the virtuality increase is temporary and it is expected that the leading parton traversing hot QCD matter experiences substantial virtuality (or mass) depletion along with energy loss \cite{Majumder:2014gda}.

The jet mass of inclusive jets and of jets in dijet events has been previously measured in high-energy pp collisions at $\sqrt{s}=7$ TeV at the LHC \cite{ATLAS:2012am,Chatrchyan:2013vbb}. Perturbative QCD predictions using higher-order matrix-elements for parton production combined with a Monte Carlo (MC) parton shower were found to be in good agreement with the data. The commonly used leading-order event generators with full shower evolution, PYTHIA \cite{Sjostrand:pythia6, Sjostrand:pythia8} and HERWIG \cite{Bahr:2008pv}, reproduce the jet mass distribution in pp collisions reasonably well in the $\pt$ region 200--600$~\GeVc$ previously studied, however they consistently under- and over-predict the data, respectively, by a slight amount.

%In this letter, measurements of the charged-jet mass are reported. Charged jets are jets clustered using only charged particles, reconstructed in the ALICE tracking system, opposed to full jets, reconstructed with both charged and neutral particles. The mass of a jet is defined as the sum of the four momenta of the jet constituents, hence it depends not only on the constituent mass, but also on their angular distance from the jet axis. 

In this letter, measurements of the charged-jet mass are reported. Charged jets are jets clustered using only charged particles, reconstructed in the ALICE tracking system, opposed to full jets, reconstructed with both charged and neutral particles. The four momentum of the jet is defined as the sum of constituent four momenta. The jet mass is calculated from the jet four-momentum, 
\begin{equation}
 M = \sqrt{E^2 - \pt^2 - \ensuremath{p_\rom{z}}^2 },
\end{equation}
where $E$ is the jet energy, $\pt$ the transverse and $p_\rom{z}$ the longitudinal momentum of the jet. % $E$, $\pt$, and $\ensuremath{p_\rom{z}}$ are calculated from the four momenta of the jets.

The measurement is performed in \PbPb{} and \pPb{} collisions and in three intervals of jet transverse momentum. Data-driven jet-by-jet background subtraction schemes are used to correct the jet mass for the contribution of the \PbPb{} underlying event. In contrast, the \pPb{} background is included in the response matrix and corrected for in the unfolding, as discussed in detail in Sec.~\ref{ssec:pbpbbkg} and \ref{ssec:pPbbkg}. The data are compared at detector level to a simulated reference without jet quenching effects. Furthermore, the measurement is corrected to particle level via a two-dimensional unfolding technique, accounting for the remaining effect of background fluctuations and detector effects. The fully corrected jet mass distribution in central \PbPb{} collisions is compared to models and to the jet mass distribution measured in \pPb{} collisions.

\section{Data sample}\label{sec:data}
The \PbPb{} collision data were recorded during the 2011 LHC \PbPb{} run at $\sNN=2.76$ TeV. This analysis used minimum-bias (MB) events, selected online by requiring a signal in the forward V0 detectors,  two arrays of scintillator tiles covering the full azimuth within $2.8<\eta<5.1$ (V0A) and $-3.7<\eta<-1.7$ (V0C). An online centrality trigger selected the 10\% most-central \PbPb{} collisions using the centrality determination as described in \cite{Abelev:2013qoq}, with 100\% efficiency for the 0--8\% centrality interval, and 60\% efficiency for the 8--10\% interval. The number of \PbPb{} events used in this analysis, after the event selection described below, is $17$ million in the 0--10\% centrality interval.

Collisions of proton and lead beams were provided by the LHC in the first months of 2013. The beam energies were 4 TeV for the proton beam and 1.58 TeV per nucleon for the lead beam, resulting in collisions at a center-of-mass energy of $\sNN = 5.02$ TeV. The nucleon--nucleon center-of-mass system moves relative to the laboratory frame with rapidity $0.465$ in the direction of the proton beam \cite{ALICE:2012xs}. In the following, $\eta$ refers to the pseudorapidity in the laboratory frame. The V0 detectors were used for online minimum bias event triggering and offline event selection. The minimum bias trigger required a signal from a charged particle in both the V0A and the V0C. The total integrated luminosity of the minimum bias event sample is 37 $\rom{\mu b}^{-1}$. In addition, events triggered by an online jet trigger using the electromagnetic calorimeter (EMCal) \cite{oai:arXiv.org:1008.0413, Adam:2015xea} were used. The online jet patch covered an area of approximately $0.2$ $\mathrm{sr}$ and required an integrated patch energy of at least 
$20$~\GeV{}.
The transverse momentum distributions of charged jets in the triggered sample was compared to the minimum bias one, showing that the trigger was fully efficient for $\ptjetch \gtrsim 60~\GeVc$. The minimum bias and triggered sample were used for unfolded $\ptjetch < 80~\GeVc$ and $\ptjetch \geq 80~\GeVc$, respectively.  The triggered sample correspond to a total integrated luminosity of 1.6 $\rom{nb}^{-1}$.

In addition to the online triggers in both collision systems, an offline selection was applied in which the online trigger was validated and remaining background events from beam--gas and electromagnetic interactions were rejected. To ensure a high tracking efficiency for all considered events, the primary vertex was required to be within 10 cm from the center of the detector along the beam axis and within 1~cm in the transverse plane \cite{Abelev:2014ffa}.

\section{Jet reconstruction and background subtraction}\label{sec:JetReco}
Jet reconstruction for both the \pPb{} and \PbPb{} analysis was performed with the \kt{} \cite{Cacciari2006} and anti-\kt{} \cite{Cacciari:2008gp} sequential recombination jet algorithms as implemented in the FastJet package \cite{Cacciari2011}. The anti-\kt{} algorithm was used for the signal jets while clusters reconstructed with the \kt{} algorithm were used to estimate the background density of the events. Jets were reconstructed using charged tracks detected in the Time Projection Chamber (TPC) \cite{Alme:2010ke} and the Inner Tracking System (ITS) \cite{Aamodt:2010aa} which cover the full azimuthal angle and pseudorapidity $|\eta|<0.9$. Jets were reconstructed using the $E$-scheme to recombine the four-vectors of the constituents, assigning the charged-pion mass for each particle. A resolution parameter, $R$, of $0.4$ was used, and the jet area was calculated by the FastJet algorithm using essentially zero momentum particles, called ghosts, with area 0.005 \cite{Cacciari:2008gn}. Jets were accepted if they were fully contained in the tracking acceptance: full azimuth and $|\eta_{\rom{
jet}}|<0.5$, to guarantee that the reconstructed jet axes were at least $R$ away from the edge of the detector acceptance.

Reconstructed tracks were accepted if their reconstructed transverse momenta exceeded 0.15 \GeVc{}, with at least 70 space points found in the TPC and at least 80\% of the geometrically accessible space-points in the TPC. Tracks were required to have at least three hits in the ITS used in the fit to ensure good track momentum resolution. To account for the azimuthally non-uniform response of the two innermost layers of the ITS, the Silicon Pixel Detector (SPD), the primary-vertex position was added to the track fit, for tracks without SPD points, in order to further improve the momentum determination of the track. The track momentum resolution was about 1\% at 1 \GeVc{} and about 3\% at 50 \GeVc \cite{Abelev:2014ffa}. Jets which contained a track with \pt{} larger than 100 \GeVc, for which the track momentum resolution exceeded 6.5\%, were rejected. The tracking efficiency in central \PbPb{} collisions was 80\% for tracks with \pt{} larger than 1 \GeVc{} and decreased to 56\% at 0.15 \GeVc. In \pPb{} collisions the tracking efficiency was 70\% for tracks with $\pt=0.15$ \GeVc{} and increased to 85\% for $\pt\geq1$ \GeVc. 

To suppress the contribution of jets consisting mainly of background particles (combinatorial jets), only jets containing a ``hard core'' were accepted. A jet was selected only if it overlapped geometrically with a jet reconstructed with only constituents with $\pt>4~\GeVc$. In the kinematic region considered, the hard core selection had similar performance as the selection used in previous works, namely demanding the jet leading track to have a transverse momentum of at least 5 \GeVc{} \cite{Abelev:2013kqa,Adam:2015ewa}.
PYTHIA pp simulations showed that applying such a selection was 100\% efficient on the jet population for charged-jet transverse momentum $\ptjetch\geq25$ \GeVc. The fluctuating background in \PbPb{} collisions affected the jet energy scale increasing the full-efficiency threshold to $\ptjetsubch=60$ \GeVc{} (where $\ptjetsubch$ is the background-subtracted $\ptjetch$, defined below in Eq.~\ref{f:ptsub}). In minimum bias \pPb{} collisions the fragmentation bias vanished for $\ptjetch\geq30$ \GeVc{}.

\subsection{Jet-by-jet background subtraction in \PbPb{} collisions}\label{ssec:pbpbbkg}
Jet measurements in \PbPb{} collisions are severely affected by the underlying event. A reconstructed jet contains particles unrelated to the hard parton shower. In this analysis the background was subtracted jet-by-jet. For this purpose, mean background densities were determined by characterizing event-by-event the contamination from soft particles unrelated to the hard jet signal. The background transverse momentum  density, $\rho$, was defined as
\begin{equation}\label{e:rho}
 \rho = \rom{median} \left\{ \frac{p_{\rom{T},i}}{A_{i}} \right\},
\end{equation}
where $i$ indicates the $i^{\rom{th}}$ \kt{} cluster in the event, $p_{\rom{T},i}$ is the transverse momentum of the cluster and $A_i$ is
its area. The two \kt{} clusters with highest transverse momentum were excluded from the calculation of the median. The average $\rho$ in the 10\% most central \PbPb{} collisions was $116$ \GeVc. 
Further details are given in \cite{Abelev:2013kqa}. 

To take into account the influence of background particles on the reconstructed jet mass, a quantity $m_{\delta,k_{\rom{T}}^{\rom{cluster}}}$ was evaluated for each \kt{} cluster following the procedure outlined in \cite{Soyez:2012hv}
\begin{equation}
 m_{\delta,k_{\rom{T}}^{\rom{cluster}}} = \sum_{j}(\sqrt{m_{\rom{j}}^{2} + p_{\rom{T,j}}^{2}c^2} - p_{\rom{T,j}}c),
\end{equation}
where the sum runs over all particles inside the \kt{} cluster, $m_{\rom{j}}$ is the mass and $p_{\rom{T,j}}$ the transverse momentum of each constituent. The background mass density is defined by
\begin{equation}
 \rho_{\rom{m}} = \rom{median} \left\{ \frac{m_{\delta,i}}{A_{i}} \right\},
\end{equation}
where the subscript $i$ indicates again the $i^{\rom{th}}$ \kt{} cluster in the event and $A_i$ is the area of the \kt{} cluster. As in the calculation of $\rho$, the two leading \kt{} clusters were excluded from the median calculation. 
For central \PbPb{} collisions, $\langle\rho_{\rom{m}}\rangle$ was found to be about $3.6$ \GeVcsq{}.

The background densities, $\rho$ and $\rho_{\rom{m}}$, were used in combination with two background subtraction techniques for jet shapes which will be described in the following:
\begin{enumerate}[label=\roman*]
 \item The area-based subtraction method corrects jet-shape observables for background or pile up effects on an event-by-event and jet-by-jet basis \cite{Soyez:2012hv}. The method is valid for any jet algorithm and infrared- and collinear- safe jet shapes. The background is characterized by $\rho$ and $\rho_{\rom{m}}$. Ghosts are added in the $\eta$-$\varphi$ plane to the event, each of them mimicking a background component in a region of area $A_\rom{g}$. 
 The shape sensitivity to pileup is determined by considering its derivatives with respect to the transverse momentum and mass of the ghosts and extrapolated by a Taylor series to zero pileup or background. 
 A complete description of the method can be found in \cite{Soyez:2012hv}.

\item The constituent subtraction method is a particle-level approach which removes or corrects jet constituents. The particle-by-particle subtraction allows to correct both the 4-momentum of the jet and its substructure. Massless ghosts are added to the event such that they cover the $\eta$-$\varphi$ plane. Each jet will therefore contain the real particles and ghosts. A distance measure is defined for each pair of a real particle $i$ and a ghost $k$:
\begin{equation}
 \Delta R_{i,k} = p_{\rom{T},i} \cdot \sqrt{(y_{i}-y_{k}^{\mathrm g})^{2} + (\varphi_{i} - \varphi_{k}^{g})^{2}},
\end{equation}
where $y$ is the rapidity and $\varphi$ the azimuthal angle.
An iterative background removal procedure starts from the particle-ghost pair with smallest distance. At each step the transverse momentum and mass of each particle and ghost are modified. The background densities $\rho$ and $\rho_{\rom{m}}$ are used to assign momentum and mass to each ghost: $p_{\rom{T}}^{g}=A_\rom{g}\rho$ and $m_{\delta}^{g}=A_\rom{g}\rho_{\rom{m}}$ where $A_\rom{g}$ is the area of each ghost. If the transverse momentum of particle $i$ is larger than the transverse momentum of the ghost, the ghost is discarded and the transverse momentum of the ghost is subtracted from the real particle. If the transverse momentum of the ghost is larger than particle $i$, the real particle is discarded and the transverse momentum of the ghost is corrected. The same procedure is applied to the mass of the particles and ghosts. All pairs are considered and the iterative procedure is terminated when the end of the list of pairs is reached. The four-momentum of the jet is recalculated with the same recombination 
scheme as used for the jet finding procedure. A complete description of the method can be found in \cite{Berta:2014eza}.
\end{enumerate}

The area-based subtraction method was used as the nominal method for the \PbPb{} analysis to correct the reconstructed jet mass for the influence of background since it is expected to induce zero bias. On the other hand, since track-by-track it is not possible to determine whether a soft particle is background or an effect of the interaction with the medium, the constituent method could potentially remove non-background particles. 

The reconstructed transverse momentum of anti-\kt{} jets, $\ptjetrawch$, is corrected according to \cite{Cacciari:2007fd}, 
\begin{equation}\label{f:ptsub}
  \ptjetsubch = \ptjetrawch - \rho \cdot A,
\end{equation}
where $A$ is the area of the jet and $\rho$ is the \pt-density of the considered event, as defined in Eq.~\ref{e:rho}. 

\subsection{Background in \pPb{} collisions}\label{ssec:pPbbkg}
In \pPb{} collisions the average $\rho$ and $\rho_{\rom{m}}$  were about $1.26$ \GeVc{} and $0.08$ \GeVcsq{}, respectively. To account for the regions of the detector without event activity, an additional correction \cite{Adam:2015hoa} was applied and the hard signal jets were excluded from the background estimate by excluding overlap of the \kt{} clusters with anti-\kt{} jets with $\ptjetch>5$ \GeVc. While the overall background contribution is significantly smaller than in \PbPb{} collisions, it was observed that the width of the mass fluctuations caused by the \pPb{} background was increased when subtracting the background on a jet-by-jet basis with respect to including it in the response.
Therefore, to minimize this effect present in sparse events and to mitigate the different sensitivities of the considered subtraction methods to fluctuations, in  \pPb{}  collisions the background  was not subtracted jet-by-jet (on an event-by-event basis), but corrected for on average in the unfolding, as explained in greater detail in Sec.~\ref{sec:JetScale}. The systematic uncertainty on this choice was assessed by subtracting the background in data with the constituent method and correcting only for the detector effects in the response (see Sec.~\ref{sec:syst}).

\section{Jet scale and resolution}\label{sec:JetScale}
%[CB: I think we should define what we mean with ``scale'', it's used also in the y-axis of Fig 2 where we show two things, mean and most probable value]
%[CB: do we need to show all the figures for the two subtraction method? MV: yes, I think we should show this. These methods are both used for this analysis and have never been used before for a measurement. Should we discuss the point of the negative masses after subtraction with the derivative method? MV: you think that is essential? not sure if it should be discussed. It's part of the method and negligible isn't it? CB: The negative mass is not negligible, the point is that we don't have good arguments. If I understand correctly from your analysis macros, you considered the negative masses in the response, since the axes starts at -20, or something, I don't remember the exact number. Instead in Fig.4 we show only the positive part. If this is the case, those negative entries can get back into the unfolded solution, so if they weren't there the result would be slightly (?) different. In Fig.~4 the constituent method doesn't allow those negative values, so if you want to describe in detail all methods at the reco level I don't see why this point is negligible.]

For the \PbPb{} analysis, the jet energy and mass response were studied by embedding simulated pp events at detector level, namely including the effects of the detector response, into real \PbPb{} events. 
The detector response was determined from a PYTHIA 6 simulation (tune A with initial state radiation parameter PARP(67) = 2.5 to fit the D0 di-jet data \cite{tuneA}) followed by a detailed particle transport using \mbox{GEANT 3} \cite{Brun:1994aa} in a detector configuration corresponding to the conditions during \PbPb{} data taking. 
% Detector effects were obtained using the generation of pp collisions with PYTHIA6 (tune A \cite{Skands:2010ak}) with a detailed detector model implemented using GEANT3 \cite{Brun1994} using the same reconstruction software settings that were used for the reconstruction of \PbPb{} events. Generator simulations utilize only prompt particles originating from the collisions ($c\tau<1$ cm). 
% In order to account for the larger tracking inefficiency due to the large occupancy in high density environment, prior to embedding the reconstructed tracks from the simulation into \PbPb{} events, an additional \pt-dependent tracking inefficiency of 2--4\% was applied \cite{Adam:2015ewa}.
Prior to embedding the reconstructed tracks from the simulation into \PbPb{} events, an additional \pt-dependent tracking inefficiency of 2--4\% was applied in order to account for the larger tracking inefficiency due to the high occupancy for large particle densities \cite{Adam:2015ewa}. The combination  of \PbPb{} and PYTHIA events will be referred to as `hybrid events'.

The same jet reconstruction procedure as in data, see Sec.~\ref{sec:JetReco}, was applied to the hybrid events, resulting in a sample of hybrid jets. 
%After embedding tracks from detector simulation into a \PbPb{} event, jets were reconstructed and subtracted for background contamination on a \mbox{jet-by-jet} basis giving a sample of hybrid jets. 
The hybrid jets were matched to the probe jets, which were obtained by reconstructing jets from only the PYTHIA events at the detector level. Not all constituents of an embedded probe jet will necessarily be found in a hybrid jet. In order to relate the hybrid to the probe jet, a matching condition was used. This required that the constituents of the hybrid jet that comes from the PYTHIA event must carry at least $50$\% of the transverse momentum of the PYTHIA jet. In the case that a hybrid jet was paired to two or more probe jets, it was matched to the probe jet with highest \pt{} and the other probe jets were considered lost, reducing the jet-finding efficiency. The efficiency in the 10\% most central events for charged jets increased from 40\% at $\ptjetsubch=10$ \GeVc{} to 100\% for $\ptjetsubch>40$ \GeVc{}.

Region-to-region fluctuations of the jet mass and \pt-scale were characterized by using the hybrid events and calculating \deltapt{} and \deltaM{}, defined as the difference between the transverse momentum or mass of the background-subtracted hybrid jet and the probe jet \cite{Abelev:2012ej}.
%Region-to-region fluctuations on the jet \pt-scale were characterized by embedding and calculating \deltapt, the difference between the transverse momentum of the background-subtracted hybrid jet and the probe jet \cite{Abelev:2012ej}. 
%Region-to-region fluctuations on the jet \pt-scale were characterized by the \deltapt-distribution. \deltapt{} was calculated as the difference between the transverse momentum of the background subtracted hybrid jet and the probe jet \cite{Abelev:2012ej}. 
%Analogously, \deltaM{} was calculated, defined as the difference between the jet mass of the background-subtracted hybrid jet and the probe jet. 
On a jet-by-jet basis a linear correlation between \deltapt{} and \deltaM{} was observed.

The jet mass distributions of hybrid jets matched to probe jets within a certain jet \pt-interval, showed on average a larger jet mass with respect to the corresponding spectrum of the probes. This offset was due to background fluctuations and limited purity and efficiency within a reconstructed jet \pt-interval, resulting in jet migration between \pt intervals.

\begin{figure}[!th]
\centering
\includegraphics[width=\linewidth]{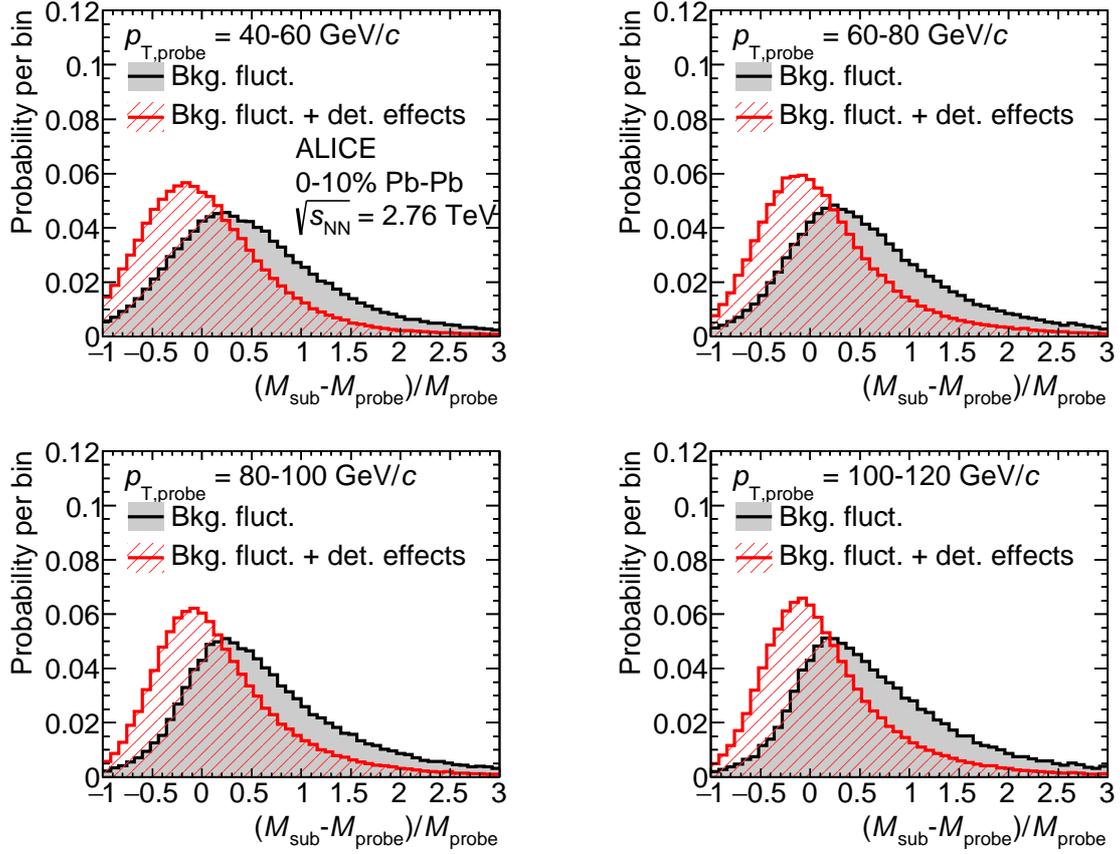}
\caption{\label{fig:DeltaMDetVsPart} Mass response using the area-based background subtraction method in the 10\% most central \PbPb{} collisions for background fluctuations only  (black histogram), compared to the full response including detector effects  (red histogram), for anti-\kt{} jets with resolution parameter $R=0.4$. $M_{\rom{sub}}$ refers to the background-subtracted reconstructed jet mass while $M_{\rom{probe}}$ is the jet mass of the embedded probe. From top left to bottom right, each panel represent a $\ptjetch$ region, 40--60, 60--80, 80--100, 100--120$~\GeVc$.
}
\end{figure}

\begin{figure}[!th]
\includegraphics[width=0.48\linewidth]{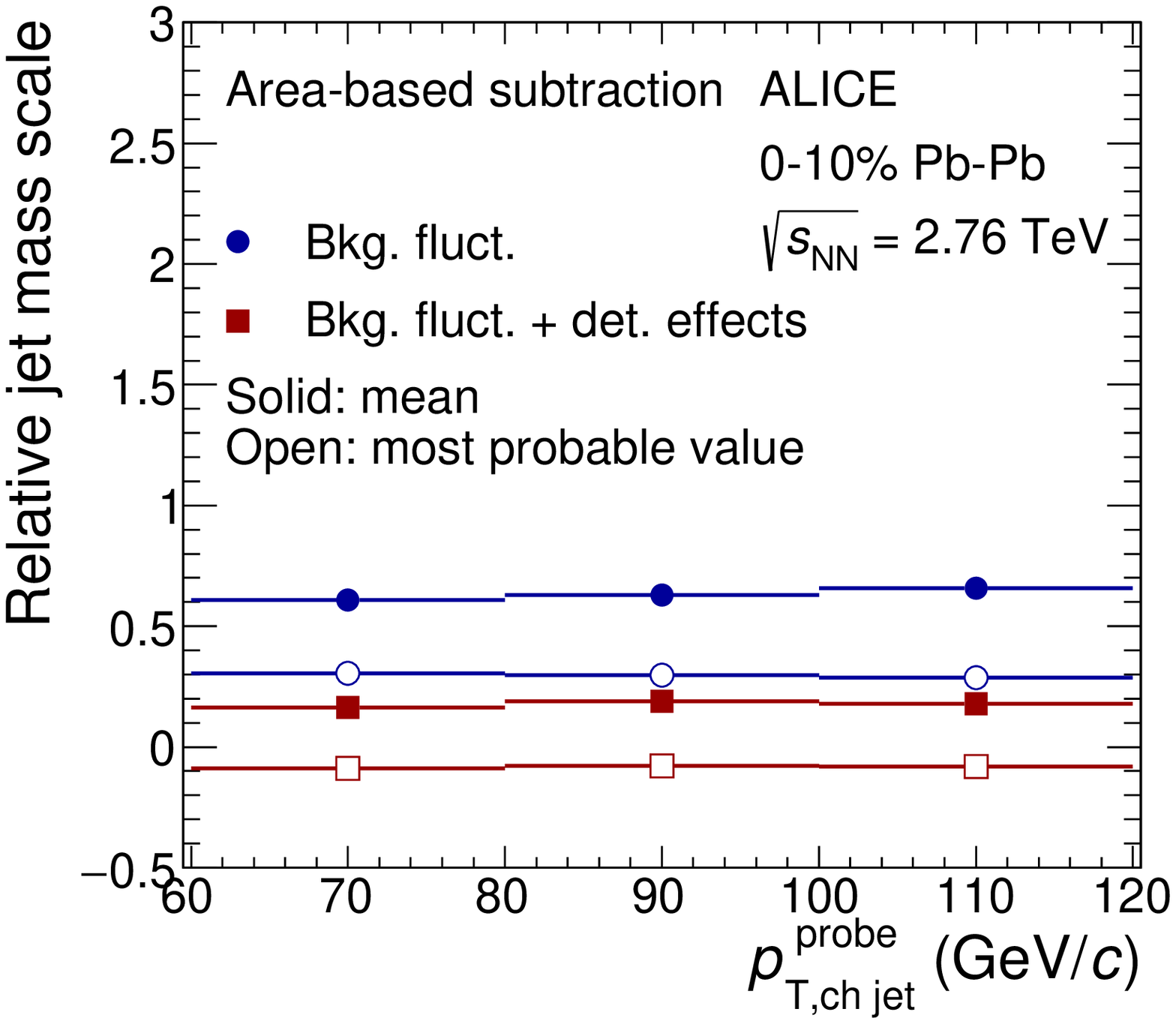}
\includegraphics[width=0.48\linewidth]{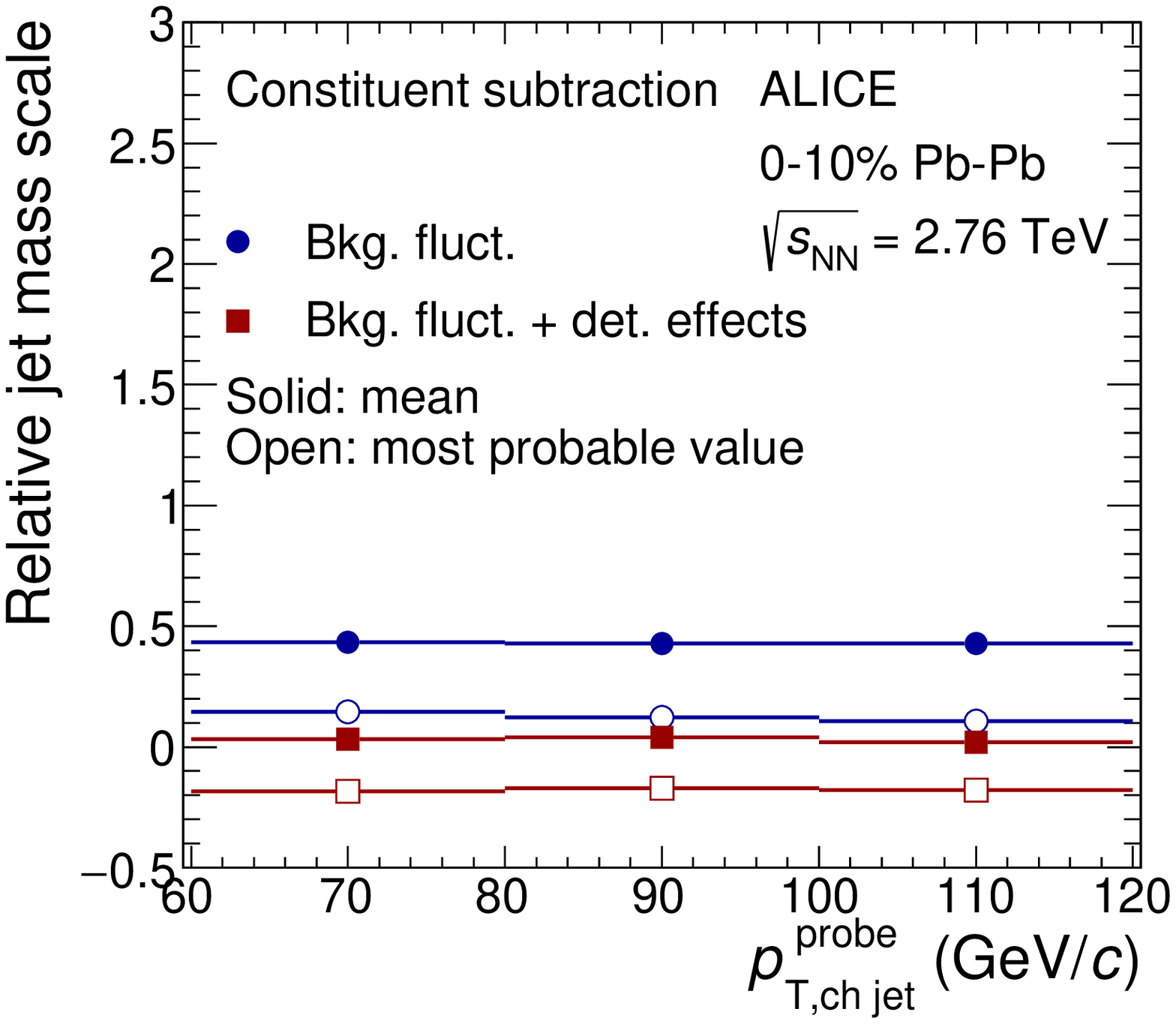}
\caption{\label{fig:JMS}Jet mass scale characterized by the relative mean and most probable value of the response. Jet mass scale is shown as a function of probe jet \pt{} for background fluctuations and the full response including detector effects, using anti-\kt{} PYTHIA jets with $R=0.4$ embedded into central \PbPb{} collisions. Left: area-based subtraction method. Right: constituent subtraction method.
}
\end{figure}

Detector effects on the jet energy and mass were investigated by matching detector level jets and particle level jets from a Pythia simulation and comparing their properties (including \PbPb{} background). The jets were matched based on distance, in a way that guarantees a one-to-one match. The constituents of the detector level jets are all assigned the pion mass, as is done for the data analysis, while the the particle mass is used for the particle level jet reconstruction. A comparison between the jet mass response due to background fluctuations and the full response, which also contains detector effects, is shown in Fig.~\ref{fig:DeltaMDetVsPart}. While background fluctuations induce a positive shift of the reconstructed jet mass, detector effects, which are dominated by the finite tracking efficiency and the mass assumption of the jet constituents, reduce the reconstructed jet mass. This was further characterized by extracting the mean and the most probable value from the distribution in Fig.~\ref{fig:DeltaMDetVsPart}, giving a measure of the relative jet mass shift. The relative mass shift is shown in Fig.~\ref{fig:JMS} for the area-based subtraction method (left) and the constituent subtraction method (right). In the kinematic range of interest, the mass shift does not exhibit a strong dependence on jet momentum. The performance of the constituent subtraction method is slightly better than for the area-based method since the constituent subtraction corrects partially for the local background fluctuations while the area-based method only corrects for the average background.

\begin{figure}[!th]
\centering
\includegraphics[width=0.5\linewidth]{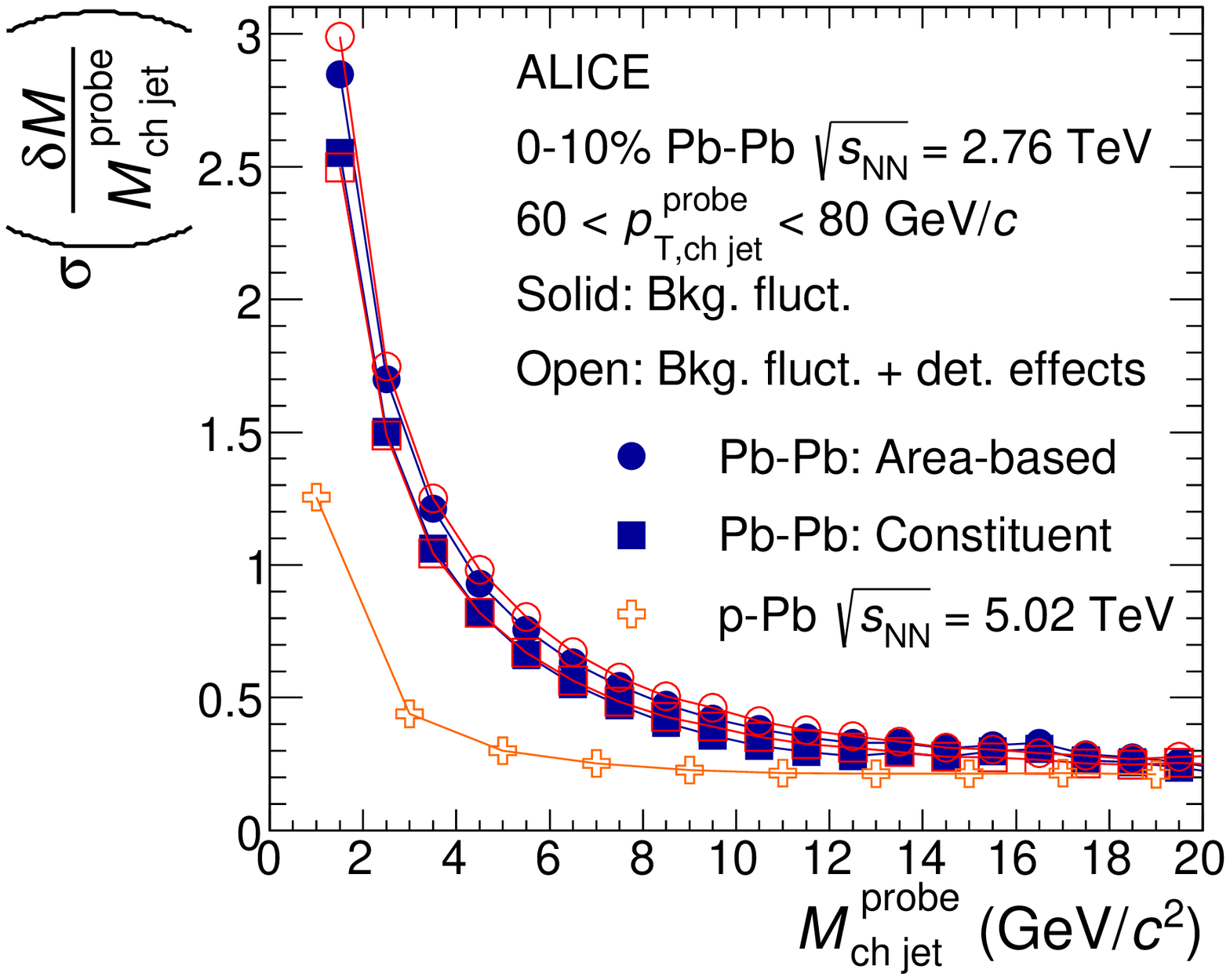}
\caption{\label{fig:JMR} Jet mass resolution as a function of true jet mass $\it{M}_{\rom{ch\;jet}}^{\rom{probe}}$ for probes with $60<{\it p}_{\rom{T,ch\;jet}}^{\rom{probe}} ~(\GeVc)<80$ \GeVc. \PbPb{}: Anti-\kt{} PYTHIA jets with $R=0.4$ embedded into central \PbPb{} collisions taking into account background fluctuations and detector effects. \pPb{}: 4-vectors corresponding to detector-level PYTHIA jets embedded into \pPb{} events. 
}
\end{figure}

Since embedding a full PYTHIA event, including the underlying event, into the sparse \pPb{} event would significantly distort the \pPb{} background estimate, the above procedure, devised with \PbPb{} collisions in mind, was modified for \pPb{} collisions. To minimize the distortion, we instead embedded \emph{single tracks}, whose 4-vectors correspond to jets reconstructed from a PYTHIA simulation (tune Perugia2011 \cite{Skands:2010ak}) at detector level, into \pPb{} events. After running FastJet on the measured events including embedded PYTHIA tracks, each resulting jet was matched with the particle level PYTHIA jet associated to the embedded track.

Figure \ref{fig:JMR} shows the jet mass resolution for \PbPb{} and \pPb{} collisions as a function of the jet mass at particle level for probe jets with $60<\ptjetch <80~\GeVc$. A strong dependence on $\it{M}_{\rom{probe}}$ is observed. The resolution for jets with a small mass is poor while for larger jet masses it improves to 25\%. Jets with small mass are very collimated and typically have a small number of constituents. The influence of the tracking inefficiency and the contamination of tracks from the background on these jets are large. For large enough \ptjetch{} ($>40$ \GeVc), jets with a small jet mass are rare and therefore the poor resolution for very collimated jets with small number of constituents is not a limiting factor in this analysis, which was restricted to jets with $\ptjetch>60$ \GeVc{} for \PbPb{} and \pPb{} collisions. For example, only about 16\% of the jets have a mass smaller than 6 \GeVcsq{} within the 60--80~\GeVc{} \ptjetch{} interval in PYTHIA.

The jet mass scale and resolution in \pPb{} collisions are dominated by tracking inefficiency, the mass assumption for the constituents, and, less strongly, by track momentum resolution. The jet mass resolution in \pPb{} collisions at small jet mass is by a factor 2 better than in \PbPb{} collisions due to the much smaller contribution of the underlying event. At large jet mass the resolution is similar for the two collision systems, 25\% for \PbPb{} and 20\% for \pPb, and mainly driven by detector effects.

%\newpage
\section{Uncorrected jet mass distributions and corrections}
\subsection{Comparison of jet mass in \PbPb{} to PYTHIA at detector level}\label{sec:embResult}
It is common use to compare uncorrected \PbPb{} results with embedded pp or PYTHIA events, including in the latter detector and background effects. We perform this comparison and then proceed with the full correction in order to compare with \pPb{} corrected results and particle-level event generators.

\begin{figure}[!th]
\centering
\includegraphics[width=0.48\linewidth]{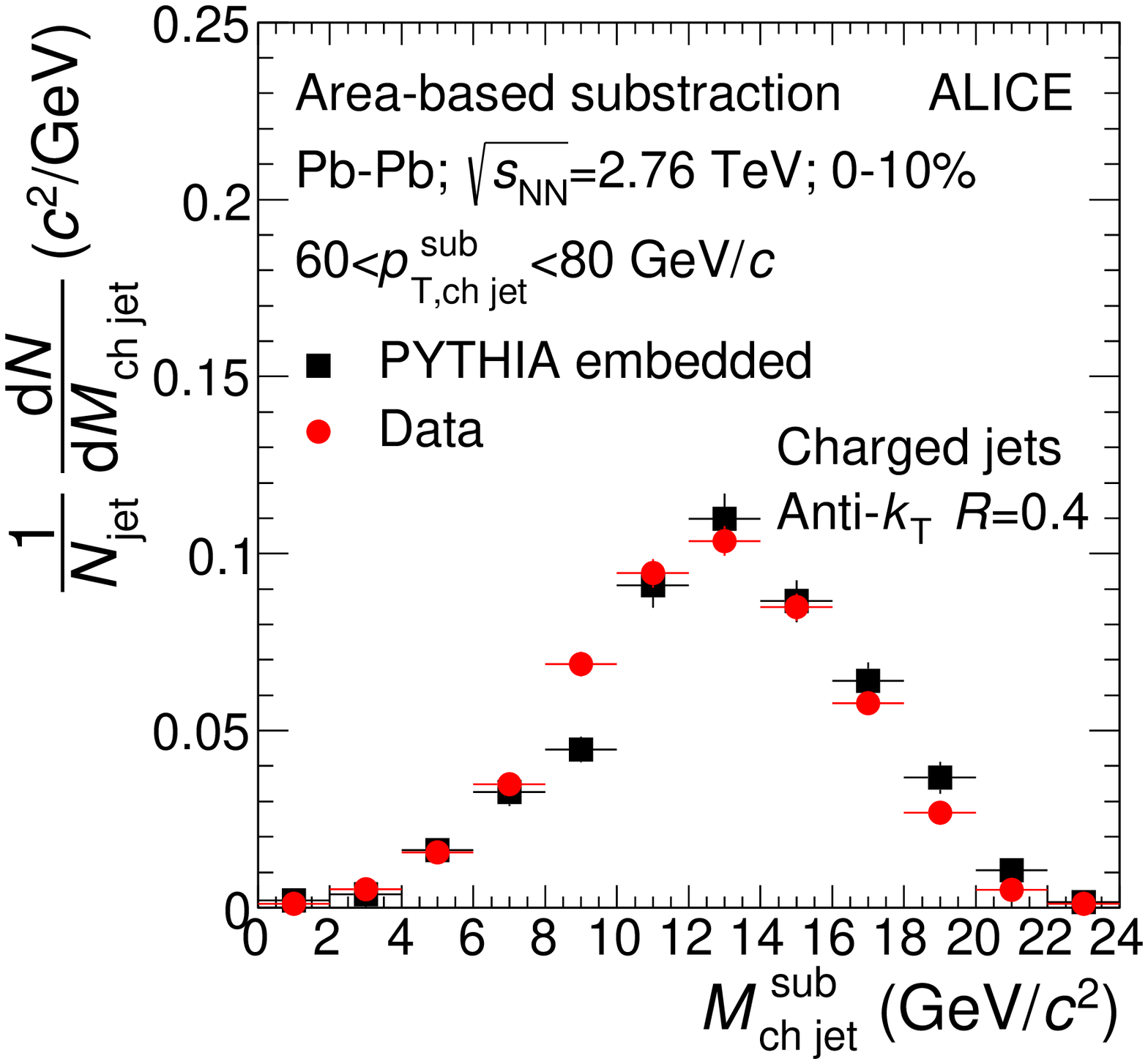}
\includegraphics[width=0.48\linewidth]{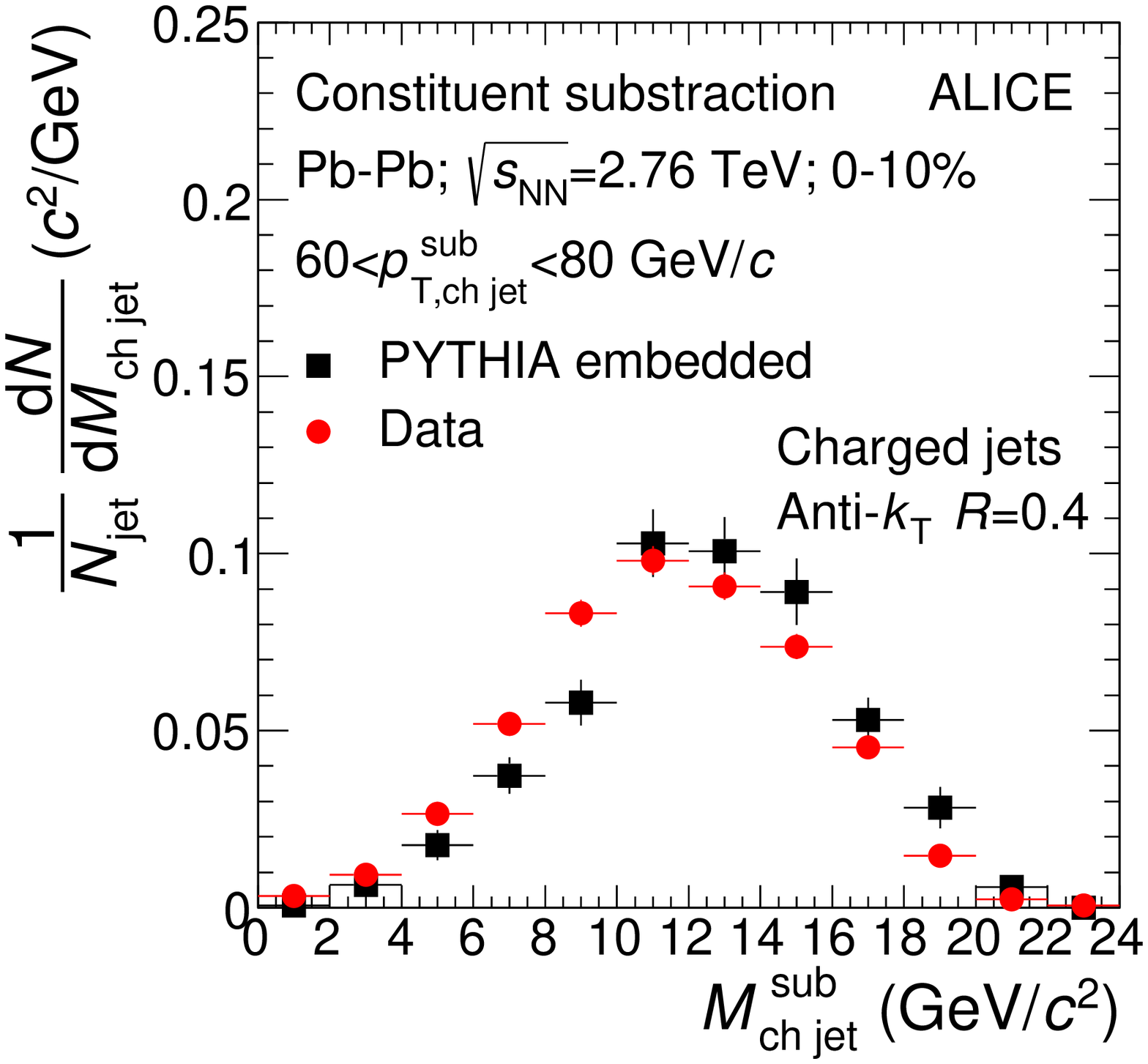}
\includegraphics[width=0.48\linewidth]{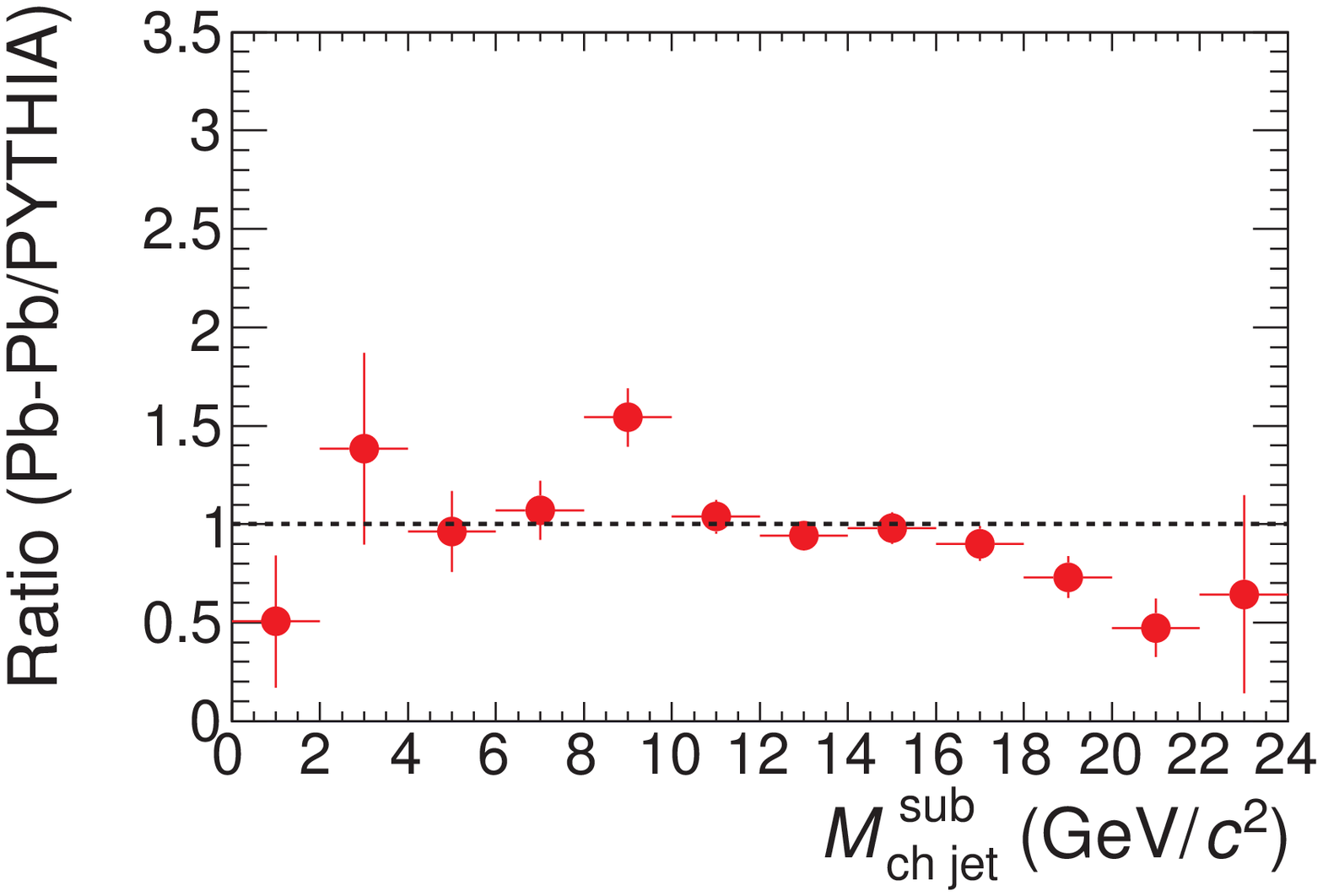}
\includegraphics[width=0.48\linewidth]{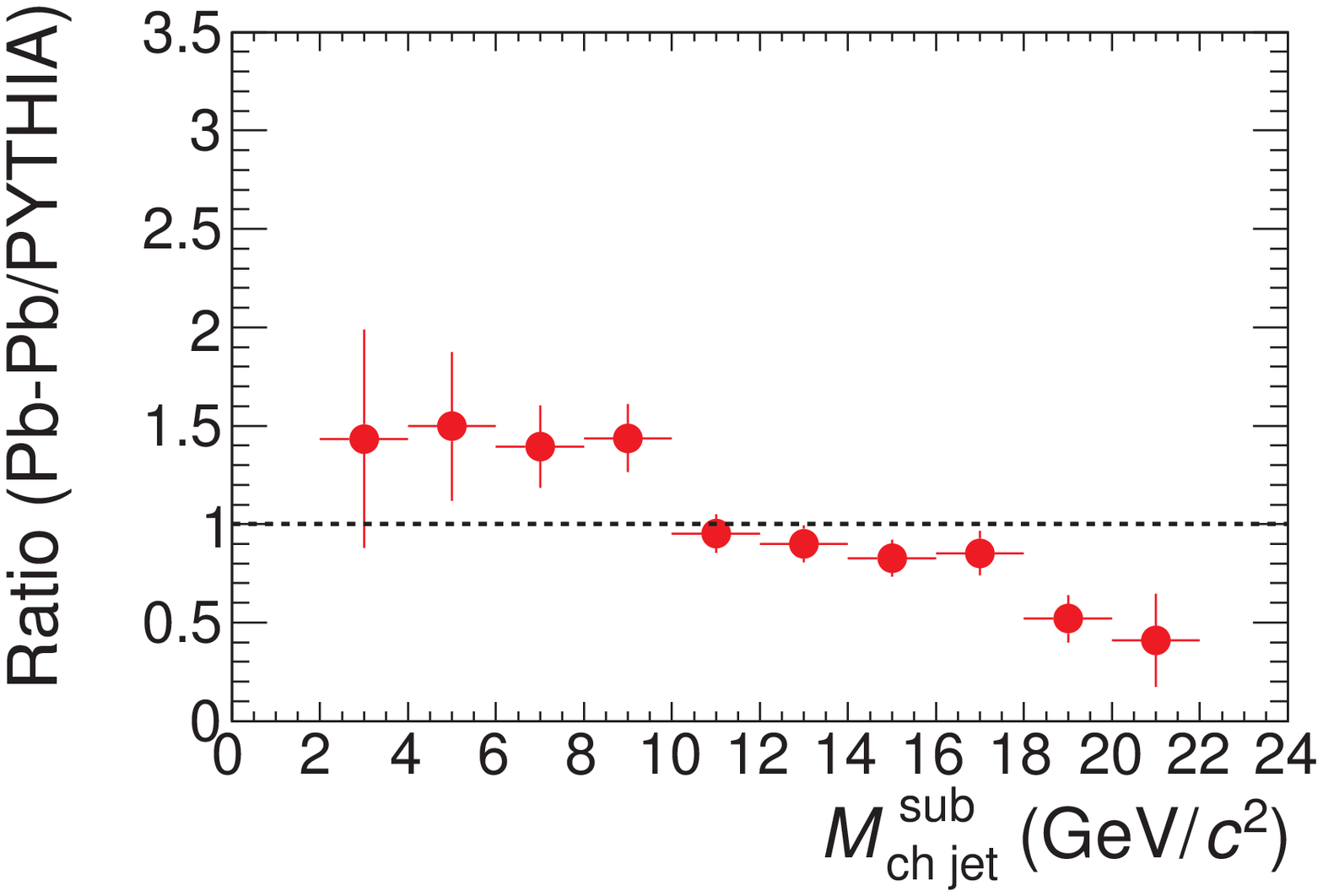}
\caption{\label{fig:JetMassCompPbPbPYTHIACent0Pt4060}Detector-level jet mass distributions in \PbPb{} data and PYTHIA (tune A) embedded into \PbPb{} collisions. Centrality: 0--10\%. Anti-\kt{} with $R=0.4$. Left: area-based background subtraction. Right: constituent background subtraction.
}
\end{figure}

\begin{figure}[!th]
\centering
\includegraphics[width=0.48\linewidth]{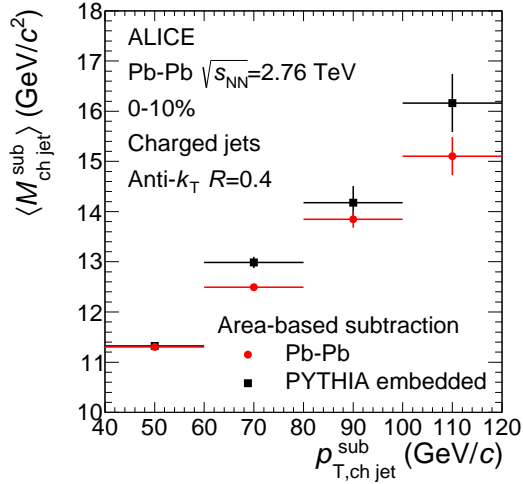}
\caption{\label{fig:MeanPbPbEmb} Comparison of the mean jet mass  in \PbPb{} collisions to detector-level embedded PYTHIA jets, for anti-\kt{} charged jets with $R=0.4$ in 10\% most central \PbPb{} collisions. Background subtraction with the area-based method.
}
\end{figure}

In this section, the jet mass distributions measured in central \PbPb{} collisions are compared to hybrid detector-level PYTHIA jets. The background was subtracted from the jet transverse momentum and mass using the area-based and constituent subtraction methods. A comparison of the jet distributions (normalized per jet) is shown in Fig.~\ref{fig:JetMassCompPbPbPYTHIACent0Pt4060}. It can be observed that the \PbPb{} and PYTHIA distributions are similar, which supports the validity of using embedded PYTHIA for the corrections as discussed in Sec.~\ref{sec:unfolding}.
The constituent method gives systematically lower jet mass than the area-based method, due to the different effect of background fluctuations for the two subtraction algorithms, see Sec. \ref{sec:JetScale}. 
The lower panels of Fig.~\ref{fig:JetMassCompPbPbPYTHIACent0Pt4060} show the ratio between \PbPb{} and PYTHIA embedded jets. 
The ratio as a function of jet mass shows that the measured distributions are very similar to the embedded PYTHIA jets, or possibly a small shift to lower mass, which is however more pronounced for the constituent background subtraction method. The hint of a shift is more pronounced in the mean jet mass, which is slightly smaller in \PbPb{} collisions than embedded PYTHIA events, as shown in Fig.~\ref{fig:MeanPbPbEmb}.  Also when comparing the corrected results with PYTHIA at particle level later in this letter, the data show a hint of a shift towards smaller masses with respect to PYTHIA when considering only statistical uncertainties.

%\newpage
\subsection{Correction of jet mass to particle level}\label{sec:unfolding}
For the correction of the jet mass measurement to particle level, a two-dimensional Bayesian unfolding technique \cite{bayeasianUnfdagostini} from RooUnfold \cite{Adye:2011gm} was used. A four-dimensional response matrix was constructed with the following axes: particle-level \ptjetch{}, detector-level \ptjetch{}, particle-level \mjetch{} and detector-level \mjetch{}. For the \PbPb{} analysis, detector-level jets were obtained by embedding detector-level PYTHIA jets into \PbPb{} events, running the jet finder and applying the background subtraction as explained in Sec. \ref{sec:JetReco}. A projection of the response on the detector level mass is shown in Fig.~\ref{fig:JetMassCompPbPbPYTHIACent0Pt4060}. 
As discussed in Sec. \ref{sec:JetScale}, the embedded detector-level jets were matched to the detector-level jets without \PbPb{} background. The latter were matched to particle-level jets in such a way to obtain a unique matching between each detector-level embedded jet and the corresponding particle-level jet. 

For the \pPb{} analysis, detector-level jets were obtained from embedding detector-level jet four-momentum vectors into \pPb{} events (see Sec. \ref{sec:JetScale}). The reconstructed embedded jets were matched with the particle-level four-momentum vectors corresponding to the detector level embedded four momenta. The four-dimensional matrix contains the smearing in jet \pt{} and mass due to background and detector effects. 

The four-dimensional response matrix was used to unfold the jet \pt{} and mass simultaneously, taking advantage of the observed strong correlation between the jet transverse momentum and mass fluctuations caused by the residual region-to-region background fluctuations, which reduces off-diagonal elements in the response matrix. The relationship between the transverse momentum and mass of the jet at particle level in the response, called the prior, is obtained from PYTHIA simulations {(tune A for \PbPb{} and Perugia 2011 for \pPb{}). A variation of this assumption was considered in the systematic uncertainties (Sec. \ref{sec:syst}).

The unfolding procedure was validated using a MC closure test by applying the correction procedure to PYTHIA embedded jets. For the signal and the response matrix, statistically independent data sets were used. The background subtracted and unfolded and true distributions agree with each other to a precision of 5\% for $\ptjetch>40$ \GeVc. The refolded distribution, obtained by convoluting the unfolded solution with the response matrix, is in agreement with the measured distribution within the statistical uncertainty.

%\newpage
\section{Systematic uncertainties}\label{sec:syst}
The systematic uncertainties for the jet mass measurement were determined by varying parameters and algorithmic choices of the measurement, corrections for detector response and background fluctuations. The main systematic uncertainties originate from the regularization of the unfolding algorithm, the background subtraction method and the uncertainty on the detector response. For the \PbPb{} analysis, also the choice of the prior, the relation between mass and $\pt$ at the particle level, used in the unfolding has an important effect. In this section the method to estimate the systematic uncertainty for each source and their magnitude in central \PbPb{} collisions  and \pPb{} collisions will be discussed.

The unfolding procedure converges after a certain number of iterations. Only relatively small variations in the results are expected when the convergence is reached. The sensitivity to the number of iterations chosen as default was estimated by varying their number over a wide range, where the convergence of the result is verified. The nominal number of iterations used for the \PbPb{} measurement is 6 and the number of iterations was varied from 3 to 10. For \pPb{} collisions the default is 3 and the number was varied between 1 and 5. Changing the number of iterations shifts the full jet mass distribution to higher or lower jet mass, resulting in an anti-correlated shape uncertainty. The relative uncertainty is largest in the tails of the jet mass distribution where it amounts to 20\% in \PbPb{} collisions and 5--20\% in \pPb{} collisions for different \pt{} ranges. In the peak region of the jet mass distributions the uncertainty does not exceed 5\% (2\%) in \PbPb{} (\pPb{}) collisions. The size of the uncertainty in the number of iterations is correlated with the statistical uncertainty and the uncertainty on the data points is correlated point-to-point.

\begin{table}
 \begin{tabular}{c | ccc | ccc}
 \hline
 Source & \multicolumn{3}{c}{\PbPb{}} & \multicolumn{3}{c}{\pPb{}}\\
 \hline
 $\ptjetch$ ($\GeVc$)      & 60--80 & 80--100 & 100--120 & 60--80 & 80--100 & 100--120 \\
 \hline
 \hline
 Prior &                         1.0\% & 3.0\% & 5.0\% & 0 & 0 & 0 \\
 Background &                    3.0\% & 3.0\% & 5.0\% & 1.0\% & 0.5\% & 1.0\%\\
 Tracking efficiency &           5.0\% & 5.0\% & 5.0\% & 3.0\% & 3.0\% & 3.0\% \\
 Unfolding (iterations, range) & 1.0\% & 3.0\% & 4.0\% & 0.5\% & 1.0\% & 4.0\%\\
 \hline
 Total &                         6.0\% & 8.0\% & 9.0\% & 3.5\% & 3.5\% & 4.5\%\\
 \hline
 \end{tabular}
\caption{\label{tab:MeanMassSyst} Systematic uncertainty in mean jet mass from different sources in the 10\% most central \PbPb{} collisions (left) and minimum-bias \pPb{} collisions (right).}
\end{table}

The prior used for the Bayesian unfolding was taken from PYTHIA simulations.
The mean jet mass as a function of uncorrected but background-subtracted jet \pt{} is 1--4\% smaller in \PbPb{} collisions than in PYTHIA simulations as shown in Fig.~\ref{fig:MeanPbPbEmb}. The second central moments of the distributions are statistically compatible indicating that the shape of the distribution is unchanged. Therefore it is reasonable to apply a shift of at maximum 4\% on the jet mass in the prior to estimate a systematic uncertainty to the measurement due to the prior choice. This results in a systematic uncertainty of 10\% around the jet mass peak, which increases gradually to 50\% in the tails.
For the \pPb{} analysis, a smearing of the mass at particle level in the response matrix was performed. The new particle level mass is extracted randomly from a Gaussian centered at the original mass with a $\sigma$ of 2\%, roughly corresponding to the maximum spread observed in the ratio of the jet mass distribution in the response at detector level and in the data. The resulting uncertainty ranges from 4\% to  6\%, with the largest value reached in the first \pt{} range.

For the jet-by-jet background subtraction in \PbPb{} collisions, the result from the area-based method was compared to the constituent subtraction. The response matrix for the methods is different since the jet mass scale differs as was shown in Fig.~\ref{fig:JMS}. The response matrix in both cases was obtained using the embedding technique presented in Sec.~\ref{sec:JetScale}. The systematic uncertainty due to the background subtraction method varies between 5\% at the center of the distribution and 30\% in the tails.

As mentioned in Sec.~\ref{ssec:pPbbkg}, in \pPb{} events the background subtraction introduces additional fluctuations due to the region-to-region fluctuations of the background, which leads to a broadening of the jet mass distribution after subtraction. It was therefore decided not to perform the subtraction event-by-event and jet-by-jet, and instead include the background in the response matrix and correct for in it the unfolding. As an extreme variation for the systematic uncertainty, the background was subtracted event-by-event in the data with the constituent method, which is less sensitive to fluctuations than the area method, and corrected only for detector effects using the PYTHIA response. The jet mass distributions corrected with the two assumptions differ by 5\% in the peak region and the difference increases gradually up to 40\% in the low-mass tail. These variations were taken as systematic uncertainties.

The uncertainty in the detector response was dominated by the uncertainty in the tracking efficiency, which was estimated by varying track quality cuts and found to be 3--4\%. The tracking efficiency in the detector simulation was varied accordingly, providing an alternative response matrix with which to repeat the unfolding. Observed differences with respect to the nominal result vary from 10\% to 40\% and 5\% to 30\% in \PbPb{} and \pPb{}, respectively, with the largest uncertainty in the tails of the distributions.

All systematic uncertainties were added in quadrature for each \mjetch{} bin. The uncertainties affect the shape of the jet mass distribution and the normalization applied causes long-range anti-correlations. 
The uncertainty on the mean jet mass as a function of \ptjetch{} was evaluated on the unfolded distribution using the variations mentioned above and  shown in Table \ref{tab:MeanMassSyst}. The total systematic uncertainty in the mean jet mass increases from 6\% for jets with $60<\ptjetch<80~\GeVc$ to 9.0\% for jets with \mbox{$100<\ptjetch<120~\GeVc$} in \PbPb{} central collisions. The systematic uncertainty in \pPb{} collisions is about two times smaller than in central \PbPb{} collisions due to the much smaller underlying event contribution.

%\newpage
\section{Results and discussion}\label{sec:result}
\subsection{Jet mass measurements in \PbPb{}  and \pPb{} collisions}\label{sec:res}

The fully unfolded jet mass distributions including all systematic uncertainties, measured in \pPb{} collisions at $\sNN=5.02$ TeV in three ranges of $\ptjetch$ between 60 and 120$~\GeVc$ are shown in Fig.~\ref{fig:pPbVsPYTHIA} and compared with PYTHIA Perugia 2011 and HERWIG EE5C \cite{Bahr:2008pv, Seymour:2013qka}. Minimum-bias triggered events were used for \mbox{$\ptjetch<80$ \GeVc{}}, while the online jet triggered event sample was used for \mbox{$\ptjetch\geq80$ \GeVc}. The agreement of data and PYTHIA is within 10--20\% for most of the $\mjetch $ range. The deviations increase for the low and high mass tail and can exceed 30--50\% for the intermediate $\ptjetch$ range. The agreement with HERWIG is slightly worse, mostly in the low mass tail of the distribution and in the highest $\ptjetch$ interval. Considering the good agreement with simulations and that the jet nuclear modification factors \ensuremath{R_{\rom{pPb}}} and \ensuremath{Q_{\rom{pPb}}} measurements show no cold nuclear matter effects \cite{Adam:2015hoa, Adam2016, CMSRjetpPbKhachatryan2016, ATLASRpPbjetAad2015392}, the \pPb{} measurement (and PYTHIA) can be used as a reference for the assessment of the hot nuclear matter effects in \PbPb{} collisions. 
 
Figure \ref{fig:DistPbPbVsPPb} shows the comparison of the jet mass distribution, normalized per jet, in central \PbPb{} collisions at $\sNN = 2.76~$TeV and the \pPb{} collision measurement. It can be observed that the jet mass distribution in \PbPb{} collisions is shifted to smaller values with respect to the measurement in \pPb{} collisions for  $\ptjetch<100$ \GeVc{}. 

Figure \ref{fig:DistPbPbVsPPbRat} shows the ratio between the jet mass distribution in the 10\% most central \PbPb{} collisions and \pPb{} collisions. The systematic uncertainties are propagated into the ratio as uncorrelated.
The center-of-mass energy at which the \PbPb{} and \pPb{} collisions were taken is different, $\sNN=2.76$ TeV for \PbPb{} and $\sNN=5.02$ TeV for \pPb{} collisions. This is expected to introduce a small difference in the jet mass distributions due to a different shape in the underlying jet \pt-spectrum and a different quark-to-gluon ratio. Therefore, the figure shows also the same ratio from particle level simulated PYTHIA pp collisions (tune Perugia 2011) at the two energies. Considering statistical uncertainties only in the ratio, a shift to lower jet masses in  \PbPb{}  is observed for $\ptjetch<100$ \GeVc{}, consistent with the PYTHIA embedded results in Sec.~\ref{sec:embResult}. Including the systematic uncertainties in our measurements, the decreasing trend of the ratio as a function of \mjetch{} is compatible between data and PYTHIA and no significant reduction in jet mass in \PbPb{} collisions is observed.

\begin{figure}[!ht]
\centering
\includegraphics[width=\linewidth]{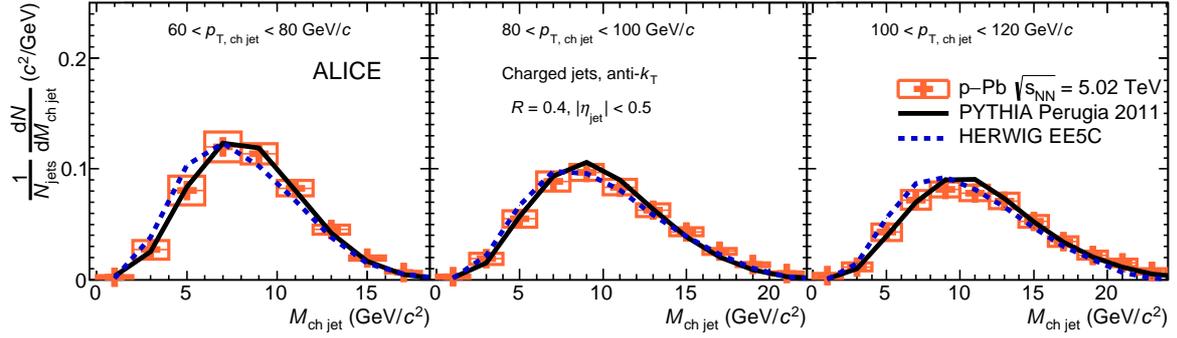}
\caption{\label{fig:pPbVsPYTHIA}Fully-corrected jet mass distribution for anti-\kt{} jets with $R=0.4$ in \pPb{} collisions, compared to PYTHIA and HERWIG simulations  for three ranges of $\ptjetch$. Statistical uncertainties in data are smaller than the markers and in the models are smaller than the line width. }
\end{figure}

\begin{figure}[!ht]
\centering
\includegraphics[width=1.0\linewidth]{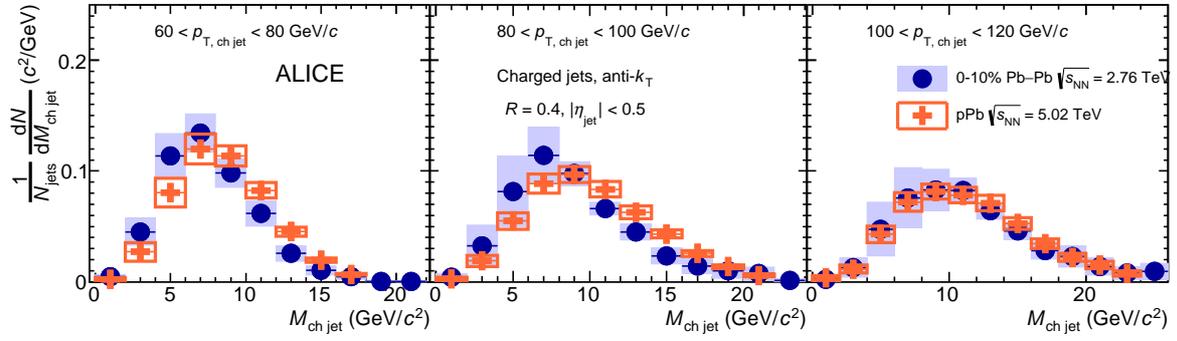}
\caption{\label{fig:DistPbPbVsPPb} Fully-corrected jet mass distribution for anti-\kt{} jets with $R=0.4$ in minimum bias \pPb{} collisions compared to central \PbPb{} collisions for three ranges of $\ptjetch$.
}
\end{figure}

\begin{figure}[!ht]
\centering
\includegraphics[width=\linewidth]{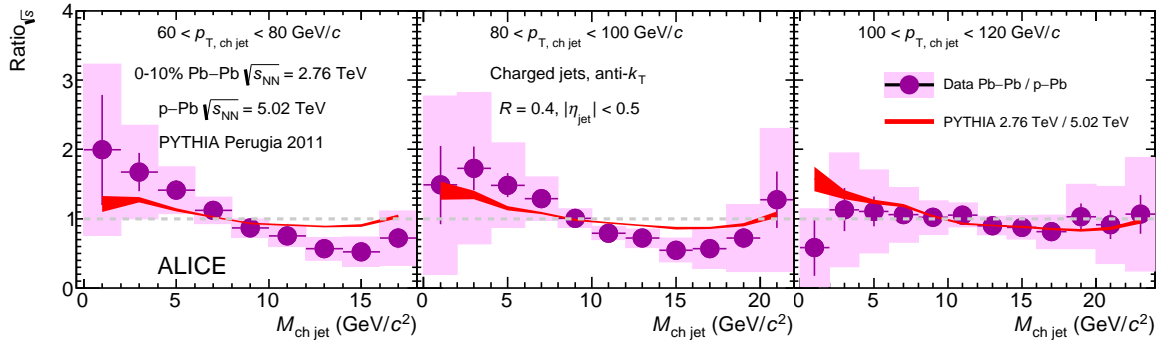}
\caption{\label{fig:DistPbPbVsPPbRat} Ratio between fully-corrected jet mass distribution for anti-\kt{} jets with $R=0.4$ in central \PbPb{} collisions and minimum bias \pPb{} collisions. The ratio is compared to the ratio of mass distributions of PYTHIA (tune Perugia 2011) at $\sqrt{s}=2.76$ TeV and $\sqrt{s}=5.02$ TeV (width of the band represents the statistical uncertainties).
}
\end{figure}

\begin{figure}[!t]
\centering
\includegraphics[width=0.48\linewidth]{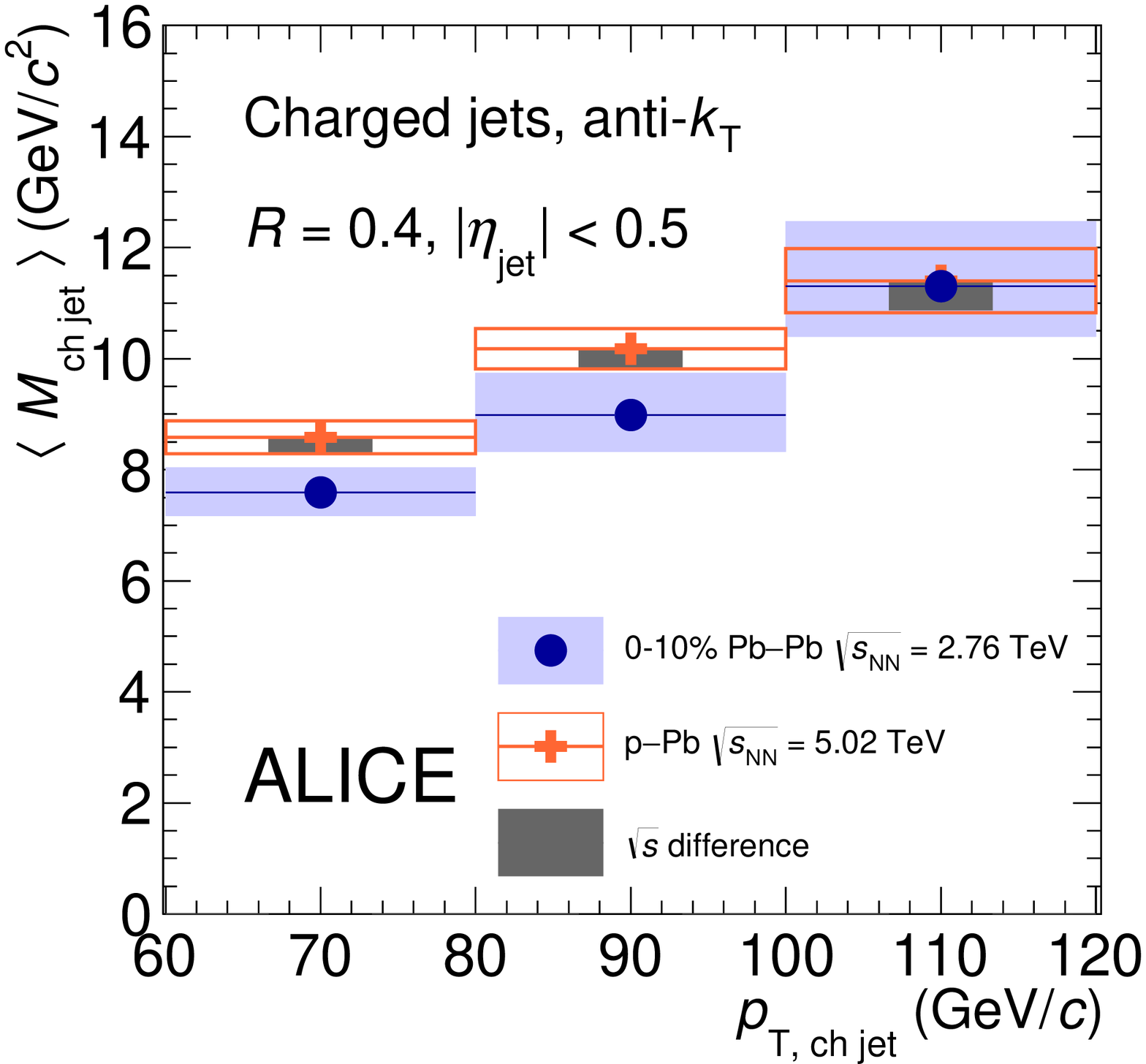}
\caption{\label{fig:MeanPbPbVsPPb} Fully-corrected mean jet mass as a function of \ptjetch{} for anti-\kt{} jets with $R=0.4$ in minimum bias \pPb{} collisions at $\sNN=5.02$ TeV compared to central \PbPb{} collisions at $\sNN=2.76$ TeV.
}
\end{figure}

The comparison of the jet mass in \PbPb{} collisions relative to \pPb{} collisions is further established by presenting the mean jet mass as a function of \ptjetch{} in Fig.~\ref{fig:MeanPbPbVsPPb}. 
The difference in the mean jet mass for the two collision energies considered is between $0.2$ and $0.5$ \GeVcsq{} in the PYTHIA simulation. This difference in the mean jet mass is indicated by a filled box attached to the \pPb{} data points in Fig.~\ref{fig:MeanPbPbVsPPb}. For the lowest \ptjetch{} range in \PbPb{} collisions the mean jet mass exhibits a reduction with respect to \pPb{} measurements, limited to about one standard deviation. For higher \ptjetch{} the mean jet mass in the two systems is compatible within systematic uncertainties.

\subsection{Model comparison and discussion}

The jet mass measurements for central \PbPb{} collisions for three \ptjetch{} intervals compared to several event generators are shown in Fig.~\ref{fig:PbPbVsModels}. PYTHIA represents the expectation without jet quenching while JEWEL \cite{Zapp:2011ek,Zapp:2012ak} and Q-PYTHIA \cite{Armesto:2009fj} (with PQM geometry \cite{Dainese:2004te})  are two models with medium-induced energy loss. 
In JEWEL each scattering of the leading parton with constituents from the medium is computed giving a microscopic description of the transport coefficient, $\hat{q}$. By default, JEWEL does not keep track of the momenta of the recoiling scattering centers (``recoil off''). This leads to a net loss of energy and momentum out of the di-jet system, and is expected to mostly affect low-$\pt$-particle production. For the jet mass measurement, low-momentum fragments are important, so JEWEL was also run in the mode in which it keeps track of the scattering centers (``recoil on''). In that mode, more soft particles are generated, some of which have very large angles with the jet and will contribute to the background estimate in the event. The JEWEL authors implemented a background subtraction in full jets by introducing ``fake" neutral constituents used for the 4-momentum subtraction. Since the pp charged jet mass distribution is reproduced by shifting the full jet mass distribution towards lower masses, the JEWEL background-subtracted charged jet mass is obtained by shifting the background-subtracted full jet mass.
Q-PYTHIA modifies the splitting functions in the PYTHIA event generator, resulting in medium-induced gluon radiation following the multiple soft scattering approximation. Both jet quenching models reproduce the suppression observed in inclusive high-\pt{} particle and jet production \cite{Zapp:2012ak,Armesto:2009fj}.

The jet mass is strongly overestimated by Q-PYTHIA due to the strong broadening of the jet profile close to the jet axis. Also JEWEL with ``recoil on'' significantly overestimates the jet mass. JEWEL ``recoil off'' underestimates the jet mass due to the large amount of out-of-cone radiation,  which does not hadronize in this mode of the generator. The vacuum expectation from PYTHIA, while slightly overestimating the jet mass for lower \ptjetch{} when considering statistical uncertainties only, is compatible  with the \PbPb{} measurement within systematic uncertainties. 
The \PbPb{} mean jet mass as a function of \ptjetch{} is compared to the event generators in Fig.~\ref{fig:MeanMassPythiaJewel}. The linear increase of the mean jet mass with jet \pt{} is expected from NLO pQCD calculations \cite{Ellis:2007ib}.

Previous jet shape and jet fragmentation function measurements clearly favor JEWEL with ``recoil on" over Q-PYTHIA \cite{Chatrchyan:2013kwa,Chatrchyan:2014ava,Zapp:2013zya,Adam2015,KunnawalkamElayavalli:2017hxo,vanLeeuwen:2015upu}. Despite the difference in the fragment distributions between Q-PYTHIA and JEWEL with ``recoil on", Fig.\ \ref{fig:PbPbVsModels} shows that both models predict a similar large increase of the jet mass, which is excluded by the measurement. JEWEL ``recoil off", which does not describe the previous measurements well because it does not include all soft radiation, gives a better description of the jet mass than JEWEL ``recoil on". The difference between the jet mass distributions in JEWEL with ``recoil on" an ``recoil off" indicates that the jet mass is sensitive to the soft fragments at large angle which are produced by hadronisation of recoil partons in the JEWEL model.

\begin{figure}[!ht]
\centering

\includegraphics[width=\linewidth]{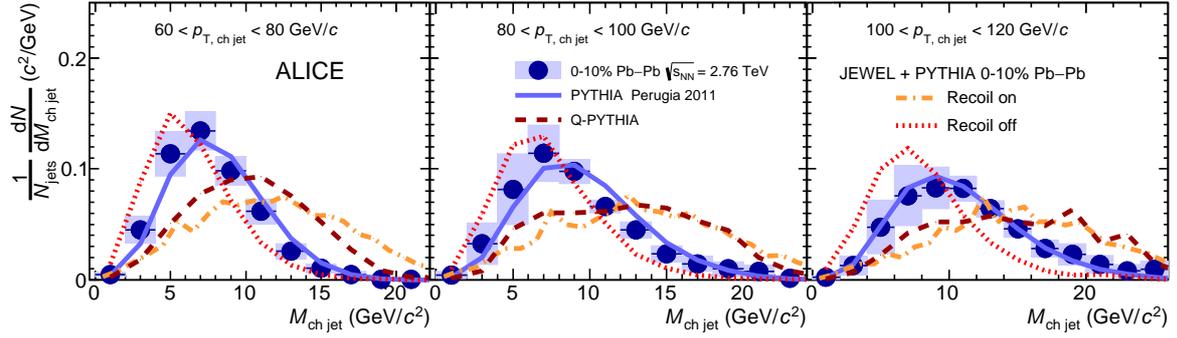}
\caption{\label{fig:PbPbVsModels} Fully-corrected jet mass distribution for anti-\kt{} jets with $R=0.4$ in the 10\% most central \PbPb{} collisions compared to PYTHIA with tune Perugia 2011 and predictions from the jet quenching event generators (JEWEL and Q-PYTHIA). Statistical uncertainties are not shown for the model calculations.
}
\end{figure}

\begin{figure}[!ht]
\centering
\includegraphics[width=0.48\linewidth]{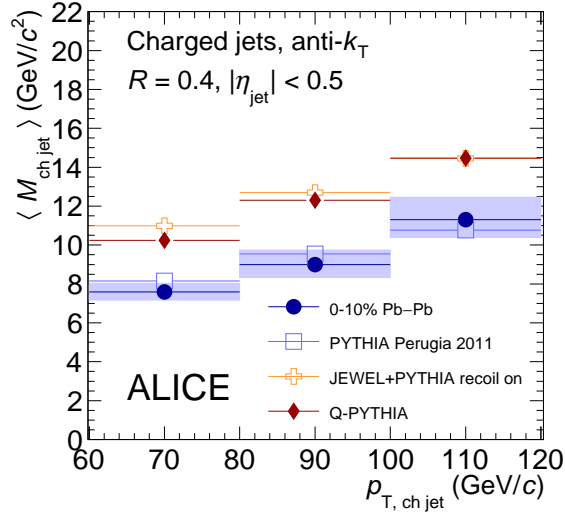}
\caption{\label{fig:MeanMassPythiaJewel} Fully-corrected mean jet mass compared to PYTHIA Perugia2011 and the jet quenching event generators (JEWEL and Q-PYTHIA) for anti-\kt{} jets with $R=0.4$ in the 10\% most central \PbPb{} collisions.
}
\end{figure}

\subsection{Summary}

The first jet mass measurement in heavy-ion collisions for charged jets ($60 < \ptjetch <120~\GeVc$) was reported and compared to \pPb{} reference measurements and models with and without quenching. 
The presented results are the first attempt to access the virtuality evolution of the hard partons in heavy-ion collisions. By constraining both energy and virtuality experimentally, differential jet mass measurements could provide further non-trivial tests for models of in-medium shower evolution.

The ratio of the jet mass distribution in central \PbPb{} collisions and minimum-bias \pPb{} collisions is compared to that in PYTHIA Perugia 2011 simulations at the two center-of-mass energies. The data ratio is compatible with the PYTHIA expectation at the two center-of-mass energies within systematic uncertainties. A hint of a difference within statistical uncertainties only in the ratio and in the mean jet mass in the lowest \ptjetch{} range is of interest to motivate further work on reducing the systematic uncertainties in order to increase the precision in jet mass measurements as well as pursuing more differential studies, for example with respect to hard fragmenting jets. 

The fully-corrected results are consistent with the observation based on detector level comparison with PYTHIA embedded jets. 
The measured jet mass in \PbPb{} collisions is not reproduced by the quenching models considered in this letter and is found to be consistent with PYTHIA vacuum expectations within systematic uncertainties. %The data lie between JEWEL with recoil off and PYTHIA. 
These results are qualitatively consistent with previous measurements of jet shapes at the LHC \cite{Chatrchyan:2013kwa,Adam2015}, which show only relatively small changes of the particle distributions in jets in  \PbPb{}  collisions compared to \pp{} collisions.
The JEWEL model with ``recoil on", which describes the existing measurements of fragment distributions in jets \cite{Chatrchyan:2013kwa,Chatrchyan:2014ava} reasonably well \cite{Zapp:2013zya,KunnawalkamElayavalli:2017hxo}, predicts a significant increase of the jet mass, contrary to what is observed in the measurement.

The observed suppression of jet yields in the presence of a dense medium, $\RAA < 1$ \cite{Aad:2014bxa}, is interpreted as due to radiated partons lost or scattered out of the jet cone. Therefore, one  reconstructs a subset of the entire parton shower within a jet with resolution parameter 0.4. In the extreme case that only the leading parton were to escape the medium, and then shower in vacuum, one would reconstruct the mass of the leading parton at the point of exit. Since also the virtuality evolution of the parton shower is modified in the presence of jet quenching, one would expect in such a scenario that the escaping (reconstructed) jets exhibit a reduced jet mass with respect to the \pp{} and \pPb{} references \cite{Majumder:2014gda}.
The data show that the jet mass is consistent within uncertainties in \PbPb{} and \pPb{} collisions within a fixed \ptjetch-interval, implying that the soft radiation outside the jet cone does not significantly alter the relation between \pt{} and the mass of the parton. 

%\input{summary}

%%%%%%%% acknowledgements
\newenvironment{acknowledgement}{\relax}{\relax}
\begin{acknowledgement}
\section*{Acknowledgements}
% Version: 2017-01-23

The ALICE Collaboration would like to thank all its engineers and technicians for their invaluable contributions to the construction of the experiment and the CERN accelerator teams for the outstanding performance of the LHC complex.
The ALICE Collaboration gratefully acknowledges the resources and support provided by all Grid centres and the Worldwide LHC Computing Grid (WLCG) collaboration.
The ALICE Collaboration acknowledges the following funding agencies for their support in building and running the ALICE detector:
A. I. Alikhanyan National Science Laboratory (Yerevan Physics Institute) Foundation (ANSL), State Committee of Science and World Federation of Scientists (WFS), Armenia;
Austrian Academy of Sciences and Nationalstiftung f\"{u}r Forschung, Technologie und Entwicklung, Austria;
Ministry of Communications and High Technologies, National Nuclear Research Center, Azerbaijan;
Conselho Nacional de Desenvolvimento Cient\'{\i}fico e Tecnol\'{o}gico (CNPq), Universidade Federal do Rio Grande do Sul (UFRGS), Financiadora de Estudos e Projetos (Finep) and Funda\c{c}\~{a}o de Amparo \`{a} Pesquisa do Estado de S\~{a}o Paulo (FAPESP), Brazil;
Ministry of Science \& Technology of China (MSTC), National Natural Science Foundation of China (NSFC) and Ministry of Education of China (MOEC) , China;
Ministry of Science, Education and Sport and Croatian Science Foundation, Croatia;
Ministry of Education, Youth and Sports of the Czech Republic, Czech Republic;
The Danish Council for Independent Research | Natural Sciences, the Carlsberg Foundation and Danish National Research Foundation (DNRF), Denmark;
Helsinki Institute of Physics (HIP), Finland;
Commissariat \`{a} l'Energie Atomique (CEA) and Institut National de Physique Nucl\'{e}aire et de Physique des Particules (IN2P3) and Centre National de la Recherche Scientifique (CNRS), France;
Bundesministerium f\"{u}r Bildung, Wissenschaft, Forschung und Technologie (BMBF) and GSI Helmholtzzentrum f\"{u}r Schwerionenforschung GmbH, Germany;
Ministry of Education, Research and Religious Affairs, Greece;
National Research, Development and Innovation Office, Hungary;
Department of Atomic Energy Government of India (DAE) and Council of Scientific and Industrial Research (CSIR), New Delhi, India;
Indonesian Institute of Science, Indonesia;
Centro Fermi - Museo Storico della Fisica e Centro Studi e Ricerche Enrico Fermi and Istituto Nazionale di Fisica Nucleare (INFN), Italy;
Institute for Innovative Science and Technology , Nagasaki Institute of Applied Science (IIST), Japan Society for the Promotion of Science (JSPS) KAKENHI and Japanese Ministry of Education, Culture, Sports, Science and Technology (MEXT), Japan;
Consejo Nacional de Ciencia (CONACYT) y Tecnolog\'{i}a, through Fondo de Cooperaci\'{o}n Internacional en Ciencia y Tecnolog\'{i}a (FONCICYT) and Direcci\'{o}n General de Asuntos del Personal Academico (DGAPA), Mexico;
Nationaal instituut voor subatomaire fysica (Nikhef), Netherlands;
The Research Council of Norway, Norway;
Commission on Science and Technology for Sustainable Development in the South (COMSATS), Pakistan;
Pontificia Universidad Cat\'{o}lica del Per\'{u}, Peru;
Ministry of Science and Higher Education and National Science Centre, Poland;
Korea Institute of Science and Technology Information and National Research Foundation of Korea (NRF), Republic of Korea;
Ministry of Education and Scientific Research, Institute of Atomic Physics and Romanian National Agency for Science, Technology and Innovation, Romania;
Joint Institute for Nuclear Research (JINR), Ministry of Education and Science of the Russian Federation and National Research Centre Kurchatov Institute, Russia;
Ministry of Education, Science, Research and Sport of the Slovak Republic, Slovakia;
National Research Foundation of South Africa, South Africa;
Centro de Aplicaciones Tecnol\'{o}gicas y Desarrollo Nuclear (CEADEN), Cubaenerg\'{\i}a, Cuba, Ministerio de Ciencia e Innovacion and Centro de Investigaciones Energ\'{e}ticas, Medioambientales y Tecnol\'{o}gicas (CIEMAT), Spain;
Swedish Research Council (VR) and Knut \& Alice Wallenberg Foundation (KAW), Sweden;
European Organization for Nuclear Research, Switzerland;
National Science and Technology Development Agency (NSDTA), Suranaree University of Technology (SUT) and Office of the Higher Education Commission under NRU project of Thailand, Thailand;
Turkish Atomic Energy Agency (TAEK), Turkey;
National Academy of  Sciences of Ukraine, Ukraine;
Science and Technology Facilities Council (STFC), United Kingdom;
National Science Foundation of the United States of America (NSF) and United States Department of Energy, Office of Nuclear Physics (DOE NP), United States of America.    %%%%%%% done by webmaster team
\end{acknowledgement}

\bibliographystyle{utphys}
\bibliography{biblio}

%%%%%%%%% appendix with author list
\newpage
\appendix
\section{The ALICE Collaboration}\label{app:collab}

% Collaboration: CERN-LHC-ALICE
% Generation Date is 2017-Jan-23

% How to use:
%%%%%%%%% appendix with author list
%\appendix
%\section{The ALICE Collaboration}
%\label{app:collab}
%\input{authors-list.tex}  %%%%%%% get the latest version before submitting

\begingroup
\small
\begin{flushleft}
S.~Acharya$^\textrm{\scriptsize 139}$,
D.~Adamov\'{a}$^\textrm{\scriptsize 87}$,
M.M.~Aggarwal$^\textrm{\scriptsize 91}$,
G.~Aglieri Rinella$^\textrm{\scriptsize 34}$,
M.~Agnello$^\textrm{\scriptsize 30}$,
N.~Agrawal$^\textrm{\scriptsize 47}$,
Z.~Ahammed$^\textrm{\scriptsize 139}$,
N.~Ahmad$^\textrm{\scriptsize 17}$,
S.U.~Ahn$^\textrm{\scriptsize 69}$,
S.~Aiola$^\textrm{\scriptsize 143}$
% ...
\renewcommand\labelenumi{\textsuperscript{\theenumi}~}

\section*{Affiliation notes}
\renewcommand\theenumi{\roman{enumi}}
\begin{Authlist}
\item \Adef{0}Deceased
\item \Adef{idp1752336}{Also at: Dipartimento DET del Politecnico di Torino, Turin, Italy}
\item \Adef{idp1771728}{Also at: Georgia State University, Atlanta, Georgia, United States}
\item \Adef{idp3971184}{Also at: M.V. Lomonosov Moscow State University, D.V. Skobeltsyn Institute of Nuclear, Physics, Moscow, Russia}
\item \Adef{idp4307472}{Also at: Department of Applied Physics, Aligarh Muslim University, Aligarh, India}
\end{Authlist}

\section*{Collaboration Institutes}
\renewcommand\theenumi{\arabic{enumi}~}

$^{1}$A.I. Alikhanyan National Science Laboratory (Yerevan Physics Institute) Foundation, Yerevan, Armenia
\\
$^{2}$Benem\'{e}rita Universidad Aut\'{o}noma de Puebla, Puebla, Mexico
\\
$^{3}$Bogolyubov Institute for Theoretical Physics, Kiev, Ukraine
\\
$^{4}$Bose Institute, Department of Physics 
and Centre for Astroparticle Physics and Space Science (CAPSS), Kolkata, India
\\
$^{5}$Budker Institute for Nuclear Physics, Novosibirsk, Russia
\\
$^{6}$California Polytechnic State University, San Luis Obispo, California, United States
\\
$^{7}$Central China Normal University, Wuhan, China
\\
$^{8}$Centre de Calcul de l'IN2P3, Villeurbanne, Lyon, France
\\
$^{9}$Centro de Aplicaciones Tecnol\'{o}gicas y Desarrollo Nuclear (CEADEN), Havana, Cuba
\\
$^{10}$Centro de Investigaciones Energ\'{e}ticas Medioambientales y Tecnol\'{o}gicas (CIEMAT), Madrid, Spain
\\
$^{11}$Centro de Investigaci\'{o}n y de Estudios Avanzados (CINVESTAV), Mexico City and M\'{e}rida, Mexico
\\
$^{12}$Centro Fermi - Museo Storico della Fisica e Centro Studi e Ricerche ``Enrico Fermi', Rome, Italy
\\
$^{13}$Chicago State University, Chicago, Illinois, United States
\\
$^{14}$China Institute of Atomic Energy, Beijing, China
\\
$^{15}$COMSATS Institute of Information Technology (CIIT), Islamabad, Pakistan
\\
$^{16}$Departamento de F\'{\i}sica de Part\'{\i}culas and IGFAE, Universidad de Santiago de Compostela, Santiago de Compostela, Spain
\\
$^{17}$Department of Physics, Aligarh Muslim University, Aligarh, India
\\
$^{18}$Department of Physics, Ohio State University, Columbus, Ohio, United States
\\
$^{19}$Department of Physics, Sejong University, Seoul, South Korea
\\
$^{20}$Department of Physics, University of Oslo, Oslo, Norway
\\
$^{21}$Department of Physics and Technology, University of Bergen, Bergen, Norway
\\
$^{22}$Dipartimento di Fisica dell'Universit\`{a} 'La Sapienza'
and Sezione INFN, Rome, Italy
\\
$^{23}$Dipartimento di Fisica dell'Universit\`{a}
and Sezione INFN, Cagliari, Italy
\\
$^{24}$Dipartimento di Fisica dell'Universit\`{a}
and Sezione INFN, Trieste, Italy
\\
$^{25}$Dipartimento di Fisica dell'Universit\`{a}
and Sezione INFN, Turin, Italy
\\
$^{26}$Dipartimento di Fisica e Astronomia dell'Universit\`{a}
and Sezione INFN, Bologna, Italy
\\
$^{27}$Dipartimento di Fisica e Astronomia dell'Universit\`{a}
and Sezione INFN, Catania, Italy
\\
$^{28}$Dipartimento di Fisica e Astronomia dell'Universit\`{a}
and Sezione INFN, Padova, Italy
\\
$^{29}$Dipartimento di Fisica `E.R.~Caianiello' dell'Universit\`{a}
and Gruppo Collegato INFN, Salerno, Italy
\\
$^{30}$Dipartimento DISAT del Politecnico and Sezione INFN, Turin, Italy
\\
$^{31}$Dipartimento di Scienze e Innovazione Tecnologica dell'Universit\`{a} del Piemonte Orientale and INFN Sezione di Torino, Alessandria, Italy
\\
$^{32}$Dipartimento Interateneo di Fisica `M.~Merlin'
and Sezione INFN, Bari, Italy
\\
$^{33}$Division of Experimental High Energy Physics, University of Lund, Lund, Sweden
\\
$^{34}$European Organization for Nuclear Research (CERN), Geneva, Switzerland
\\
$^{35}$Excellence Cluster Universe, Technische Universit\"{a}t M\"{u}nchen, Munich, Germany
\\
$^{36}$Faculty of Engineering, Bergen University College, Bergen, Norway
\\
$^{37}$Faculty of Mathematics, Physics and Informatics, Comenius University, Bratislava, Slovakia
\\
$^{38}$Faculty of Nuclear Sciences and Physical Engineering, Czech Technical University in Prague, Prague, Czech Republic
\\
$^{39}$Faculty of Science, P.J.~\v{S}af\'{a}rik University, Ko\v{s}ice, Slovakia
\\
$^{40}$Faculty of Technology, Buskerud and Vestfold University College, Tonsberg, Norway
\\
$^{41}$Frankfurt Institute for Advanced Studies, Johann Wolfgang Goethe-Universit\"{a}t Frankfurt, Frankfurt, Germany
\\
$^{42}$Gangneung-Wonju National University, Gangneung, South Korea
\\
$^{43}$Gauhati University, Department of Physics, Guwahati, India
\\
$^{44}$Helmholtz-Institut f\"{u}r Strahlen- und Kernphysik, Rheinische Friedrich-Wilhelms-Universit\"{a}t Bonn, Bonn, Germany
\\
$^{45}$Helsinki Institute of Physics (HIP), Helsinki, Finland
\\
$^{46}$Hiroshima University, Hiroshima, Japan
\\
$^{47}$Indian Institute of Technology Bombay (IIT), Mumbai, India
\\
$^{48}$Indian Institute of Technology Indore, Indore, India
\\
$^{49}$Indonesian Institute of Sciences, Jakarta, Indonesia
\\
$^{50}$Inha University, Incheon, South Korea
\\
$^{51}$Institut de Physique Nucl\'eaire d'Orsay (IPNO), Universit\'e Paris-Sud, CNRS-IN2P3, Orsay, France
\\
$^{52}$Institute for Nuclear Research, Academy of Sciences, Moscow, Russia
\\
$^{53}$Institute for Subatomic Physics of Utrecht University, Utrecht, Netherlands
\\
$^{54}$Institute for Theoretical and Experimental Physics, Moscow, Russia
\\
$^{55}$Institute of Experimental Physics, Slovak Academy of Sciences, Ko\v{s}ice, Slovakia
\\
$^{56}$Institute of Physics, Academy of Sciences of the Czech Republic, Prague, Czech Republic
\\
$^{57}$Institute of Physics, Bhubaneswar, India
\\
$^{58}$Institute of Space Science (ISS), Bucharest, Romania
\\
$^{59}$Institut f\"{u}r Informatik, Johann Wolfgang Goethe-Universit\"{a}t Frankfurt, Frankfurt, Germany
\\
$^{60}$Institut f\"{u}r Kernphysik, Johann Wolfgang Goethe-Universit\"{a}t Frankfurt, Frankfurt, Germany
\\
$^{61}$Institut f\"{u}r Kernphysik, Westf\"{a}lische Wilhelms-Universit\"{a}t M\"{u}nster, M\"{u}nster, Germany
\\
$^{62}$Instituto de Ciencias Nucleares, Universidad Nacional Aut\'{o}noma de M\'{e}xico, Mexico City, Mexico
\\
$^{63}$Instituto de F\'{i}sica, Universidade Federal do Rio Grande do Sul (UFRGS), Porto Alegre, Brazil
\\
$^{64}$Instituto de F\'{\i}sica, Universidad Nacional Aut\'{o}noma de M\'{e}xico, Mexico City, Mexico
\\
$^{65}$IRFU, CEA, Universit\'{e} Paris-Saclay, F-91191 Gif-sur-Yvette, France, Saclay, France
\\
$^{66}$iThemba LABS, National Research Foundation, Somerset West, South Africa
\\
$^{67}$Joint Institute for Nuclear Research (JINR), Dubna, Russia
\\
$^{68}$Konkuk University, Seoul, South Korea
\\
$^{69}$Korea Institute of Science and Technology Information, Daejeon, South Korea
\\
$^{70}$KTO Karatay University, Konya, Turkey
\\
$^{71}$Laboratoire de Physique Corpusculaire (LPC), Clermont Universit\'{e}, Universit\'{e} Blaise Pascal, CNRS--IN2P3, Clermont-Ferrand, France
\\
$^{72}$Laboratoire de Physique Subatomique et de Cosmologie, Universit\'{e} Grenoble-Alpes, CNRS-IN2P3, Grenoble, France
\\
$^{73}$Laboratori Nazionali di Frascati, INFN, Frascati, Italy
\\
$^{74}$Laboratori Nazionali di Legnaro, INFN, Legnaro, Italy
\\
$^{75}$Lawrence Berkeley National Laboratory, Berkeley, California, United States
\\
$^{76}$Moscow Engineering Physics Institute, Moscow, Russia
\\
$^{77}$Nagasaki Institute of Applied Science, Nagasaki, Japan
\\
$^{78}$National and Kapodistrian University of Athens, Physics Department, Athens, Greece, Athens, Greece
\\
$^{79}$National Centre for Nuclear Studies, Warsaw, Poland
\\
$^{80}$National Institute for Physics and Nuclear Engineering, Bucharest, Romania
\\
$^{81}$National Institute of Science Education and Research, Bhubaneswar, India
\\
$^{82}$National Nuclear Research Center, Baku, Azerbaijan
\\
$^{83}$National Research Centre Kurchatov Institute, Moscow, Russia
\\
$^{84}$Niels Bohr Institute, University of Copenhagen, Copenhagen, Denmark
\\
$^{85}$Nikhef, Nationaal instituut voor subatomaire fysica, Amsterdam, Netherlands
\\
$^{86}$Nuclear Physics Group, STFC Daresbury Laboratory, Daresbury, United Kingdom
\\
$^{87}$Nuclear Physics Institute, Academy of Sciences of the Czech Republic, \v{R}e\v{z} u Prahy, Czech Republic
\\
$^{88}$Oak Ridge National Laboratory, Oak Ridge, Tennessee, United States
\\
$^{89}$Petersburg Nuclear Physics Institute, Gatchina, Russia
\\
$^{90}$Physics Department, Creighton University, Omaha, Nebraska, United States
\\
$^{91}$Physics Department, Panjab University, Chandigarh, India
\\
$^{92}$Physics Department, University of Cape Town, Cape Town, South Africa
\\
$^{93}$Physics Department, University of Jammu, Jammu, India
\\
$^{94}$Physics Department, University of Rajasthan, Jaipur, India
\\
$^{95}$Physikalisches Institut, Eberhard Karls Universit\"{a}t T\"{u}bingen, T\"{u}bingen, Germany
\\
$^{96}$Physikalisches Institut, Ruprecht-Karls-Universit\"{a}t Heidelberg, Heidelberg, Germany
\\
$^{97}$Physik Department, Technische Universit\"{a}t M\"{u}nchen, Munich, Germany
\\
$^{98}$Purdue University, West Lafayette, Indiana, United States
\\
$^{99}$Pusan National University, Pusan, South Korea
\\
$^{100}$Research Division and ExtreMe Matter Institute EMMI, GSI Helmholtzzentrum f\"ur Schwerionenforschung GmbH, Darmstadt, Germany
\\
$^{101}$Rudjer Bo\v{s}kovi\'{c} Institute, Zagreb, Croatia
\\
$^{102}$Russian Federal Nuclear Center (VNIIEF), Sarov, Russia
\\
$^{103}$Saha Institute of Nuclear Physics, Kolkata, India
\\
$^{104}$School of Physics and Astronomy, University of Birmingham, Birmingham, United Kingdom
\\
$^{105}$Secci\'{o}n F\'{\i}sica, Departamento de Ciencias, Pontificia Universidad Cat\'{o}lica del Per\'{u}, Lima, Peru
\\
$^{106}$Sezione INFN, Bari, Italy
\\
$^{107}$Sezione INFN, Bologna, Italy
\\
$^{108}$Sezione INFN, Cagliari, Italy
\\
$^{109}$Sezione INFN, Catania, Italy
\\
$^{110}$Sezione INFN, Padova, Italy
\\
$^{111}$Sezione INFN, Rome, Italy
\\
$^{112}$Sezione INFN, Trieste, Italy
\\
$^{113}$Sezione INFN, Turin, Italy
\\
$^{114}$SSC IHEP of NRC Kurchatov institute, Protvino, Russia
\\
$^{115}$Stefan Meyer Institut f\"{u}r Subatomare Physik (SMI), Vienna, Austria
\\
$^{116}$SUBATECH, IMT Atlantique, Universit\'{e} de Nantes, CNRS-IN2P3, Nantes, France
\\
$^{117}$Suranaree University of Technology, Nakhon Ratchasima, Thailand
\\
$^{118}$Technical University of Ko\v{s}ice, Ko\v{s}ice, Slovakia
\\
$^{119}$Technical University of Split FESB, Split, Croatia
\\
$^{120}$The Henryk Niewodniczanski Institute of Nuclear Physics, Polish Academy of Sciences, Cracow, Poland
\\
$^{121}$The University of Texas at Austin, Physics Department, Austin, Texas, United States
\\
$^{122}$Universidad Aut\'{o}noma de Sinaloa, Culiac\'{a}n, Mexico
\\
$^{123}$Universidade de S\~{a}o Paulo (USP), S\~{a}o Paulo, Brazil
\\
$^{124}$Universidade Estadual de Campinas (UNICAMP), Campinas, Brazil
\\
$^{125}$Universidade Federal do ABC, Santo Andre, Brazil
\\
$^{126}$University of Houston, Houston, Texas, United States
\\
$^{127}$University of Jyv\"{a}skyl\"{a}, Jyv\"{a}skyl\"{a}, Finland
\\
$^{128}$University of Liverpool, Liverpool, United Kingdom
\\
$^{129}$University of Tennessee, Knoxville, Tennessee, United States
\\
$^{130}$University of the Witwatersrand, Johannesburg, South Africa
\\
$^{131}$University of Tokyo, Tokyo, Japan
\\
$^{132}$University of Tsukuba, Tsukuba, Japan
\\
$^{133}$University of Zagreb, Zagreb, Croatia
\\
$^{134}$Universit\'{e} de Lyon, Universit\'{e} Lyon 1, CNRS/IN2P3, IPN-Lyon, Villeurbanne, Lyon, France
\\
$^{135}$Universit\'{e} de Strasbourg, CNRS, IPHC UMR 7178, F-67000 Strasbourg, France, Strasbourg, France
\\
$^{136}$Universit\`{a} degli Studi di Pavia, Pavia, Italy
\\
$^{137}$Universit\`{a} di Brescia, Brescia, Italy
\\
$^{138}$V.~Fock Institute for Physics, St. Petersburg State University, St. Petersburg, Russia
\\
$^{139}$Variable Energy Cyclotron Centre, Kolkata, India
\\
$^{140}$Warsaw University of Technology, Warsaw, Poland
\\
$^{141}$Wayne State University, Detroit, Michigan, United States
\\
$^{142}$Wigner Research Centre for Physics, Hungarian Academy of Sciences, Budapest, Hungary
\\
$^{143}$Yale University, New Haven, Connecticut, United States
\\
$^{144}$Yonsei University, Seoul, South Korea
\\
$^{145}$Zentrum f\"{u}r Technologietransfer und Telekommunikation (ZTT), Fachhochschule Worms, Worms, Germany
\endgroup

  %%%%%%% get the latest version before submitting

\end{document}